\documentclass[aps,12pt]{revtex4} 
\usepackage[dvips]{epsfig} 
\usepackage{amsmath} 
\usepackage{amssymb} 

\parskip=\medskipamount
 
\renewcommand{\arraystretch}{1.2} 

\def\be{\begin{equation}}
\def\ee{\end{equation}} 
\def\disp{\displaystyle}

\begin{document} 

\title{On metric structure of ultrametric spaces} 
\author{S.K. Nechaev$^{1,2}$, O.A. Vasilyev$^{2}$}
\affiliation{$^1$LPTMS, Universit\'e Paris Sud, 91405 Orsay Cedex, France \\ 
$^2$Landau Institute for Theoretical Physics, 117334, Moscow, Russia} 

\date{\today} 

\begin{abstract} 

In our work we have reconsidered the old problem of diffusion at the boundary 
of ultrametric tree from a "number theoretic" point of view. Namely, we use 
the modular functions (in particular, the Dedekind $\eta$--function) to 
construct the "continuous" analog of the Cayley tree isometrically embedded in 
the Poincare upper half-plane. Later we work with this continuous Cayley tree 
as with a standard function of a complex variable. In the frameworks of our 
approach the results of Ogielsky and Stein on dynamics on ultrametric spaces 
are reproduced semi-analytically/semi-numerically. The speculation on the new 
"geometrical" interpretation of replica $n\to 0$ limit is proposed. 

\end{abstract} 

\maketitle 

\tableofcontents 

\section{Introduction} 
\label{sect:1} 

Let us begin with an obvious and almost tautological statement: any regular 
tree has an ultrametric structure. Recall that ultrametricity of the space 
${\cal H}$ implies the "strong triangle inequality" meaning that the distances 
$r_{AB}$, $r_{BC}$ and $r_{AC}$ between any three points $A,B,C$ in the space 
${\cal H}$ satisfy the condition $r_{AC}\le\max\left\{r_{AB}, r_{BC}\right\}$ 
(see, for example, \cite{rammal}). Appearing in mathematical literature in 
connection with $p$--adic analysis (see, for review \cite{freund,volovich}), 
ultrametric spaces became very popular in physical community because of their 
importance for spin glasses---see, for review \cite{mezard}, and references 
therein.

The famous replica symmetry breaking (RSB) scheme \cite{parisi} is ultimately 
connected to the ultrametric structure of the phase space of many disordered 
systems possessing the spin--glass behavior. Since invention of RSB, many 
authors, both physicists and mathematicians, attempted to adapt the $p$--adic 
analysis for physical needs mainly trying to elucidate and justify the replica 
$n\to 0$ limit. However to our point of view the interpretation of spin--glass 
problems in terms of $p$--adic language does not converge well. As an 
exception one has to mention recent interesting contributions to that subject 
\cite{avet1,sourl,avet2}, where the application of $p$--adic Fourier transform 
allows significantly simplify the solution to the problem of diffusion on 
ultrametric tree. One can hope that the continuation and generalization of 
these works would release deeper penetration of $p$--adic analysis into 
physics of disordered systems with ultrametric phase spaces, such as spin 
glasses, neural networks and disordered heteropolymers (see \cite{avet2,avet3} 
for the detailed list of corresponding references).

In what follows we shall always keep in mind the $p+1$--branching Cayley tree 
as an example of an ultrametric space. Despite extremely simple topological 
structure of a tree, one cannot operate in this space as in usual space with 
an Euclidean metric because the number of degrees of freedom for the 
$p+1$--branching Cayley tree grows exponentially with the size of a tree. For 
some problems, such as, for example, the branching process on Cayley trees and 
tree--like graphs it is not sufficient to deal only with the "distance" 
measured in number of generations between two points on the graph, but it is 
ultimately necessary to know the absolute values of coordinates of points on 
the Cayley tree. The main difficulty concerns the encoding of the Cayley tree 
vertices. This problem becomes very cumbersome because for the tree we do not 
have any transparent and convenient "coordinate system", like a D--dimensional 
grid in a D--dimensional Euclidean space. One of possible ways to resolve the 
addressed problem consists in using the $p$--adic analysis \cite{dom1,dom2}. 
The vertices of the graph ${\cal C}$, i.e. of the $p+1$--branching Cayley 
tree, admit the natural parameterizations by the $p$--adic numbers what 
enables to develop the whole machinery like $p$--adic Fourier transform etc. 
This way has been exploited in papers \cite{avet1,sourl,avet2}. 

Another possibility, described in the present paper consists in the following. 
Instead of working with the ultrametric discrete graph ${\cal C}$, one can 
embed this graph in the metric space ${\cal H}$ preserving all ultrametric 
properties of ${\cal C}$. Let us mention that any image of a regular Cayley 
tree refers indirectly to the isometric embedding of a tree into the complex 
plane. Indeed, the Cayley tree is usually drawn in the picture in such a way 
that any new generation of vertices (counted from the root point) is smaller 
than the previous generation in geometric progression. Taking advantage of the 
isometric embedding, one can: 

\begin{itemize} 
\item[---] naturally parameterize the vertices of the Cayley graph ${\cal C}$ 
by ordinary complex variable $z=x+iy$ in the complex plane $z$ without using 
any ingredients of a $p$--adic analysis; 
\item[---] construct a continuous analog of a Cayley tree, i.e. a continuous 
space ${\cal H}$ with ultrametric properties borrowed from the initial Cayley 
tree. 
\end{itemize} 

The paper is organized as follows. In the section \ref{sect:2} we describe the 
way of isometric embedding of the regular 3--branching Cayley tree ${\cal C}$ 
into the complex upper half-plane $z=x+iy$ as well as we discuss the 
ultrametric properties of the resulting space ${\cal H}$. The problem of 
diffusion on the boundary of ultrametric tree ${\cal C}$ embedded into the 
space ${\cal H}$ is considered in the section \ref{sect:3}. The results are 
summarized in the conclusion, where we express also some conjectures 
concerning the possible geometrical interpretation of replica $n\to 0$ limit. 

\section{Ultrametric structure of isometric Cayley trees} 
\label{sect:2} 

It is well known that any regular Cayley tree, as an exponentially growing 
structure, cannot be isometrically embedded in an Euclidean plane. Recall that 
the embedding of a Cayley tree ${\cal C}$ into the metric space is called 
"isometric" if ${\cal C}$ covers that space, preserving all angles and 
distances. For example, the rectangular lattice isometrically covers the 
Euclidean plane ${\cal E}=\{x,y\}$ with the flat metric $ds^2=dx^2+dy^2$. In 
the same way the Cayley tree ${\cal C}$ isometrically covers the surface of 
constant negative curvature (the Lobachevsky plane) ${\cal H}$. One of 
possible representations of ${\cal H}$, known as a Poincar\'e model, is the 
upper half-plane ${\rm Im}\,z>0$ of the complex plane $z=x+iy$ endowed with 
the metric $ds^2=\frac{dx^2+dy^2} {y^2}$ of constant negative curvature. 

In this Section we are aimed to construct the "continuous" analog of the 
standard 3--branching Cayley tree by means of modular functions and analyse 
the structure of the barriers separating the neighboring valleys. Due to 
specific number--theoretic properties of modular functions these barriers are 
ultrametrically organized. 

\subsection{Continuous analog of isometric Cayley tree} 

To be precise, let us begin with the explicit description of the standard 
recursive construction which allows encoding of all vertices of the 
3--branching Cayley tree ${\cal C}$ isometrically covering the surface of the 
constant negative curvature ${\cal H}=\{z|\,{\rm Im}\,z>0\}$. Recall that the 
3--branching Cayley tree is the Cayley graph of the group $\Lambda$ which has 
the free product structure: $\Lambda\sim \mathbb{Z}_2\otimes 
\mathbb{Z}_2\otimes \mathbb{Z}_2$ (where $\mathbb{Z}_2$ is the cyclic group of 
2nd order). The matrix representation of the generators $h_1,h_2,h_3$ of the 
group $\Lambda$ is well known (see, for example \cite{terras}): 
\be 
h_1=\left(\begin{array}{cc} 1 & -\frac{2}{\sqrt{3}} \\ 0 & -1 \end{array} 
\right),\; h_2=\left(\begin{array}{cc} 1 & \frac{2}{\sqrt{3}} \\ 0 & -1 
\end{array} \right),\; h_3=\left(\begin{array}{cc} 0 & \frac{1}{\sqrt{3}} \\ 
\sqrt{3} & 0 \end{array} \right) \label{eq:1} 
\ee 
For our purposes it is convenient to take a framing consisting of the 
composition of the standard fractional--linear transform and the complex 
conjugacy. Namely, denoting by $\bar z$ the complex conjugate of $z$, we 
consider the following action in ${\cal H}$: 
\be 
\left(\begin{array}{cc} a & b \\ c & d \end{array} \right):\; z\to\frac{a\bar 
z + b}{c\bar z + d} 
\label{eq:1a} 
\ee 
We raise the Cayley tree ${\cal C}$ in ${\cal H}$ as follows. Take the set of 
generators $\{h_1,h_2,h_3\}$ of the group ${\Lambda}$. Choose the point 
$(x_0,iy_0)=(0,i)$ as the root of the tree. 

\begin{figure}[ht] 
\begin{center} 
\epsfig{file=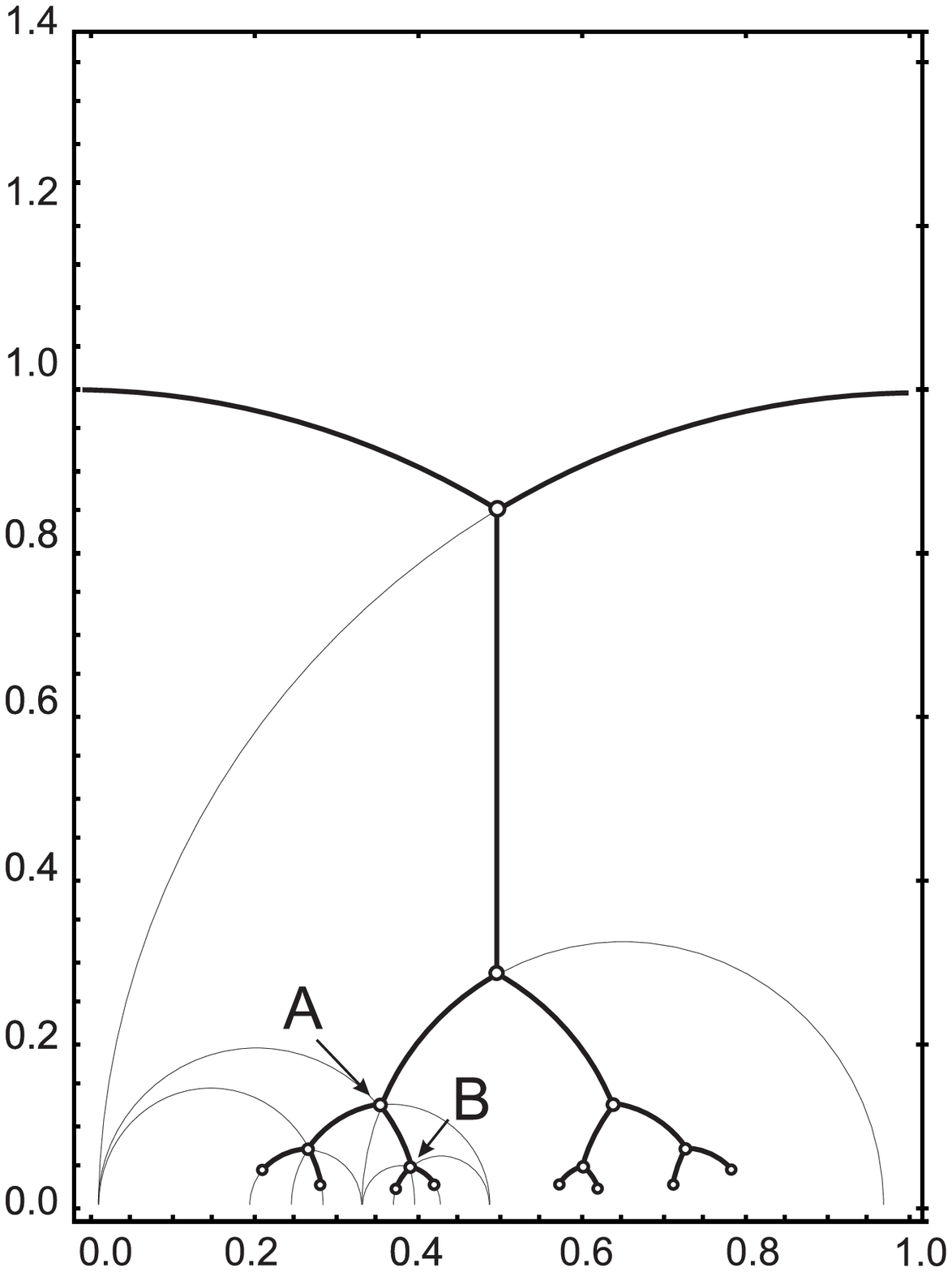,width=6cm} 
\epsfig{file=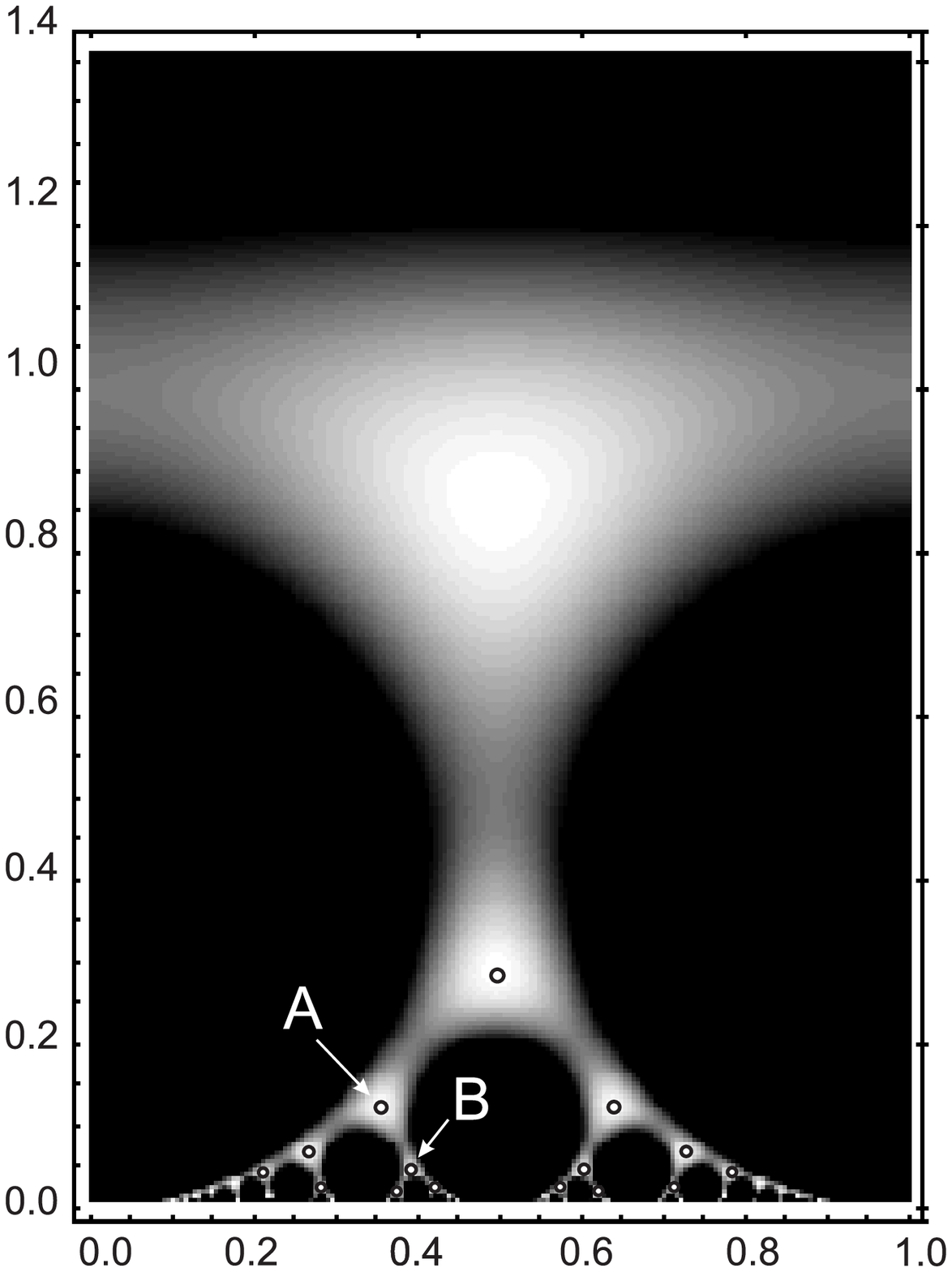,width=6cm} 
\end{center} 
\caption{Left: 3--branching Cayley tree isometrically embedded in Poincar\'e 
hyperbolic upper half-plane ${\cal H}$. Right: Density plot of the function 
$f(z)$ (see the text) in the rectangle $\{0\le {\rm Re}\,z\le 1,\; 0.01\le 
{\rm Im}\,z\le 1.4\}$.} 
\label{fig:1} 
\end{figure}

Any vertex in the generation $n$ from the root point of the tree ${\cal C}$ is 
associated with an element $\disp M_n=\prod_{k=1}^n h_{\alpha_k}$ of the group 
$\Lambda$ (where $\alpha_k\in\{1,2,3\}$ for any $k$). All vertices are 
parameterized by the complex coordinates $z_n=M_{n} \Big((-1)^n\,i\Big)$ in 
${\cal H}$. For example, the coordinates of the points $z_A$ and $z_B$ in 
fig.\ref{fig:1} we can get the following way. Multiply the generators $h_i$ 
along the trajectory in the backway order, i.e. {\it from} the points $z_A$ 
and $z_B$ {\it to} the root point $i$. Hence we arrive at 
$$ 
\begin{array}{l} 
M_2^{(A)}\equiv\left(\begin{array}{cc} a & b \\ c & d 
\end{array}\right)= h_3\,h_2= \left(\begin{array}{cc} 0 & \frac{1}{\sqrt{3}} 
\\ \sqrt{3} & 0 \end{array} \right) \left(\begin{array}{cc} 1 & 
\frac{2}{\sqrt{3}} \\ 0 & -1 \end{array} \right)= \left(\begin{array}{cc} 0 & 
-\frac{1}{\sqrt{3}} \\ \sqrt{3} & 2 \end{array} \right) \medskip \\ M_3^{(B)} 
\equiv \left(\begin{array}{cc} a & b \\ c & d \end{array}\right)= 
h_3\,h_2\,h_3= \left(\begin{array}{cc} 0 & \frac{1}{\sqrt{3}} \\ \sqrt{3} & 0 
\end{array} \right) \left(\begin{array}{cc} 1 & \frac{2}{\sqrt{3}} \\ 0 & -1 
\end{array} \right) \left(\begin{array}{cc} 0 & \frac{1}{\sqrt{3}} \\ \sqrt{3} 
& 0 \end{array} \right)= \left(\begin{array}{cc} -1 & 0 \\ 2\sqrt{3} & 1 
\end{array} \right) \end{array} $$ Using (\ref{eq:1a}) and the matrices 
$M_2^{(A)}$ and $M_3^{(B)}$, we get $$ \begin{array}{l} \disp z_A 
=\frac{0\times i - \frac{1}{\sqrt{3}}}{\sqrt{3}\times i + 2} = 
-\frac{2}{7\sqrt{3}} + \frac{i}{7} \medskip \\ \disp z_B= \frac{(-1)\times 
(-i) + 0}{2\sqrt{3}\times (-i) + 1} = -\frac{2\sqrt{3}}{13} + \frac{i}{13} 
\end{array} 
$$ 
To establish the connection with the forthcoming constructions, we shall make 
the simple linear transform of $z: z\to \frac{\sqrt{3}}{2} z+\frac{1}{2}$. The 
coordinates of the points $A$ and $B$ are shown in Fig.\ref{fig:1} after such 
transform.

However such recursive construction has, to our point of view, the crucial 
lack: one cannot automatically encode all vertices of the tree introducing a 
sort of a "coordinate system" as, say, in the Euclidean plane. Actually, we 
know that any vertex of a rectangular grid in the Euclidean plane has the 
coordinates $(a\,m,\;b\,n)$, where $\{m,n\}\in Z$ and $(a,b)$ are the length 
and the width of the elementary cell of the lattice. Our desire is to 
construct somewhat similar for homogeneous trees. To be more specific, we are 
aimed to find an analytic function $f(z)$ defined in ${\rm Im}\,z>0$ whose all 
zeros give all coordinates of the vertices of the 3--branching homogeneous 
Cayley tree ${\cal C}$ isometrically embedded into ${\cal H}(z|{\rm 
Im}\,z>0)$. In the rest of this Section we describe the construction of the 
corresponding function $f(z|{\rm Im}\,z>0)$ and discuss its properties in 
connection with the geometry of ultrametric spaces. The forthcoming 
construction is based on the properties of the Dedekind $\eta$--function. 
Recall the standard definition of the $\eta(z)$--function (see, for instance 
\cite{chand}): 
\be 
\eta(z)=e^{\pi i z/12}\prod_{k=0}^{\infty}(1-e^{2\pi i k z}); \qquad {\rm 
Im}\,z>0 
\label{eq:2} 
\ee 
It is well known \cite{chand} that the Dedekind $\eta$--function is connected 
to the elliptic Jacobi $\vartheta$--functions by the following relation: 
\be 
\vartheta_1'(0,e^{\pi i z})=\eta^3(z) 
\label{eq:3} 
\ee 
where 
\be 
\vartheta_1'(0,e^{\pi i z})\equiv \frac{d\vartheta_1(u,e^{\pi i 
z})}{du}\bigg|_{u=0}= 2e^{\pi i z/4}\sum_{n=0}^{\infty}(-1)^n (2n+1)e^{\pi i 
n(n+1)z} 
\label{eq:4} 
\ee 
Now we are in position to formulate the central assertion of the paper. Define 
the function $f(z)$ as follows: 
\be 
f(z)=C^{-1}\,|\eta(z)|\, ({\rm Im}\,z)^{1/4} 
\label{eq:5} 
\ee 
where $C$ is the normalization constant, 
\be 
C=\left|\eta\left(\frac{1}{2}+i\frac{\sqrt{3}}{2}\right)\right|\, 
\left(\frac{\sqrt{3}}{2}\right)^{1/4}=0.77230184... 
\label{eq:5a} 
\ee 
The normalization constant $C$ is chosen to set the maximal value of the 
function $f(z)$ to 1: $0<f(z)\le 1$ for any $z$ in the upper half-plane ${\rm 
Im}\,z>0$. It can be proved that the function $f(z)$ has the following 
remarkable properties: 

\begin{enumerate} 
\item The function $f(z)-1$ has zeros at the points $z_{\rm cen}$ (and only at 
these points) -- the centers of zero--angled triangles tesselating the 
Poincar\'e hyperbolic upper half-plane ${\cal H}(z|{\rm Im}\,z>0)$; 
\item The function $f(z)$ has local maxima at all the points $z_{\rm cen}$ and 
only at these points. 
\end{enumerate} 

\begin{figure}[ht] 
\begin{center} 
\epsfig{file=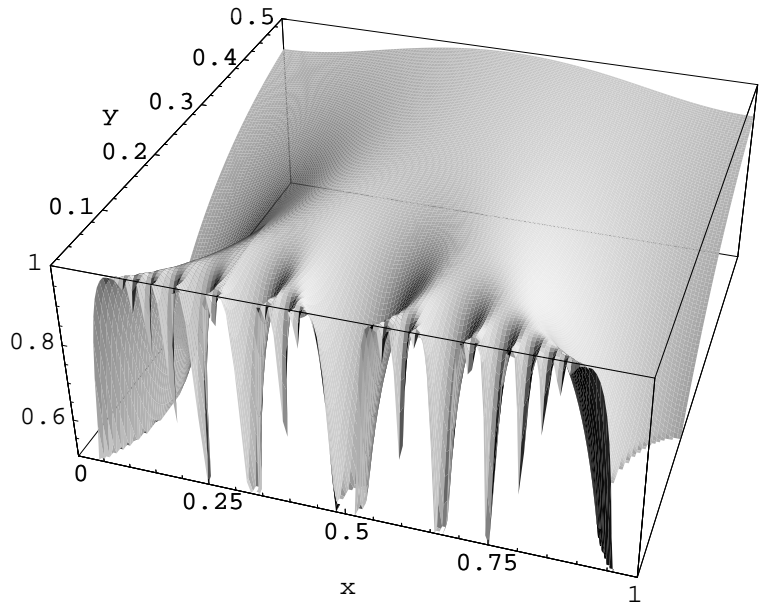,width=10cm} 
\end{center} 
\caption{Relief of the function $f(z)$ in the rectangle $\{0\le {\rm Re}\,z\le 
1,\; 0.04\le {\rm Im}\,z\le 1.4\}$.} 
\label{fig:2} 
\end{figure}

All the solutions of the equation $f(z)-1=0$ define all the coordinates of the 
3--branching Cayley tree isometrically embedded into the upper half-plane 
${\cal H}(z|{\rm Im}\,z>0)$. The corresponding density plot of the function 
$f(z)$ in the region $\{0\le {\rm Re}\,z \le 1,\; 0.04\le {\rm Im}\,z\le 
1.4\}$ is shown in fig.\ref{fig:2}. It is noteworthy that the function 
$Z(z)=\left[C\,f(z)\right]^{-2}$ is the partition function of a free bosonic 
field on a torus \cite{franc}. The function $Z(z)$ has been extensively 
studied in the conformal field theory. In particular it can be easily checked 
that the function $Z(z)$ is invariant with respect to the action of the 
modular group $PSL(2,\mathbb{Z})$, namely: 
\be 
\left\{\begin{array}{l} Z(z) = Z(z+1) \medskip \\ \disp Z(z) = 
Z\left(-\frac{1}{z}\right) \end{array}\right. 
\label{eq:6} 
\ee 
Later on we shall address to these self--dual properties. The proof of our 
assertion (\ref{eq:5}) implies the check that the function $f(z)$ is invariant 
with respect to the conformal transform $z^{(1)}(z)$ of the Poincar\'e 
half-plane to itself 
\be 
z^{(1)}(z)=\frac{z-z_0}{z\bar{z}_0-1} 
\label{eq:7} 
\ee 
where $z_0$ is the coordinate of any center of zero--angled triangle in the 
hyperbolic Poincar\'e upper half--plane obtained by successive transformations 
from the initial one. Hence, it is necessary and sufficient to show that the 
values of the function $f(z)$ at nearest centers of circular triangles are 
equal and reach the maximal value 1: 
\be 
f\left(z=\frac{1}{2}+i\frac{\sqrt{3}}{2}\right)= f\left(z=\pm 
1+\frac{1}{2}+i\frac{\sqrt{3}}{2}\right)= 
f\left(z=\frac{1}{2}+i\frac{\sqrt{3}}{6}\right)=1 \label{eq:8} 
\ee 
Then taking $z_0= \left\{\left(\frac{3}{2}+i\frac{\sqrt{3}}{2}\right); 
\left(-\frac{1}{2}+ i\frac{\sqrt{3}}{2}\right); 
\left(\frac{1}{2}+i\frac{\sqrt{3}}{6}\right) \right\}$ and performing the 
conformal transform (\ref{eq:7}) we move the centers of the first generation 
of zero--angled triangles to the new centers (second generation) located at 
$z^{(1)}$. Now we can repeat recursively the construction, i.e. find the new 
coordinates of the centers of the third generation of zero--angled triangles, 
$z^{(2)}(z^{(1)})$, and compute the function $f(z)$ at these points, then we 
perform the conformal transform $z^{(3)} (z^{(2)})$ and so on... Let us use 
now the following well--known properties of Jacobi elliptic 
$\vartheta$--functions \cite{chand} 
\be 
\eta\left(\frac{pz+r}{qz+s}\right)=\omega \sqrt{rz+s}\;\eta(z) 
\label{eq:8a} 
\ee 
where ${\rm Im}\,z>0$, $\{p,q,r,s\}\in Z$; $ps-qr=1$ and $\omega$ is some 
24--power root of unity, which depends on the coefficients $p,q,r,s$ but does 
not depend on $z$. Using (\ref{eq:8a}) we can write: 
\be 
\left\{\begin{array}{ll} \eta(z+k)=\eta(z); & \quad k=\pm 1,\pm 2,\pm 3,... 
\medskip \\ 
\eta\left(\frac{1}{2}+\frac{i}{2\sqrt{\lambda}}\right)\lambda^{-1/4}= 
\eta\left(\frac{1}{2}+\frac{i\sqrt{\lambda}}{2}\right); & \quad \lambda>0 
\end{array}\right. 
\label{eq:9} 
\ee 
Now we can easily check the relations (\ref{eq:8}). In turn, (\ref{eq:6}) can 
be regarded as a particular case of eq.(\ref{eq:8a}). The fact that the 
function $f(z)$ has local maxima at the points $z_{\rm cen}$ can be verified 
straightforwardly by differentiating $f(z)$ with respect to $z$ at the points 
$z_{\rm cen}$. The 3D relief of the function $f(z)$ is shown in figure 
\ref{fig:2}. The tree--like structure of hills separated by the valleys is 
clearly seen. 

Let us note the important duality relation for the Dedekind function which 
establishes the mapping $y\leftrightarrow \frac{1}{y}$ (where $y={\rm Im}\,z$; 
$y>0$). For all $\{p,q,r,s\}\in Z$ such that $ps-qr=1$ the following equation 
is satisfied:
\be 
\left|\eta\left(\left\{\frac{p}{q}\right\}+i\,y\right)\right|= 
\left|\eta\left(\left\{\frac{s}{q}\right\}+\frac{i}{q^2\,y}\right)\right|\; 
\frac{1}{\sqrt{q\,y}} \label{eq:9a} 
\ee 
where by $\left\{\frac{p}{q}\right\}$ and $\left\{\frac{s}{q}\right\}$ we 
denote the fractional parts of the corresponding quotients. The relation 
(\ref{eq:9a}) shall be used in the forthcoming sections for practical purposes 
of numerical computations. In the next chapter we discuss the ultrametric 
organization of corresponding barriers. We regard the 3D relief constructed on 
the basis of the function $f(z)$ (see fig.\ref{fig:2}) as the continuous 
analog of the Cayley tree isometrically embedded in ${\cal H}(z|{\rm 
Im}\,z>0)$.

\subsection{Ultrametric structure of barriers in the standard RSB scheme and 
in the tree--like metric space} 

\begin{figure}[ht] 
\begin{center} 
\epsfig{file=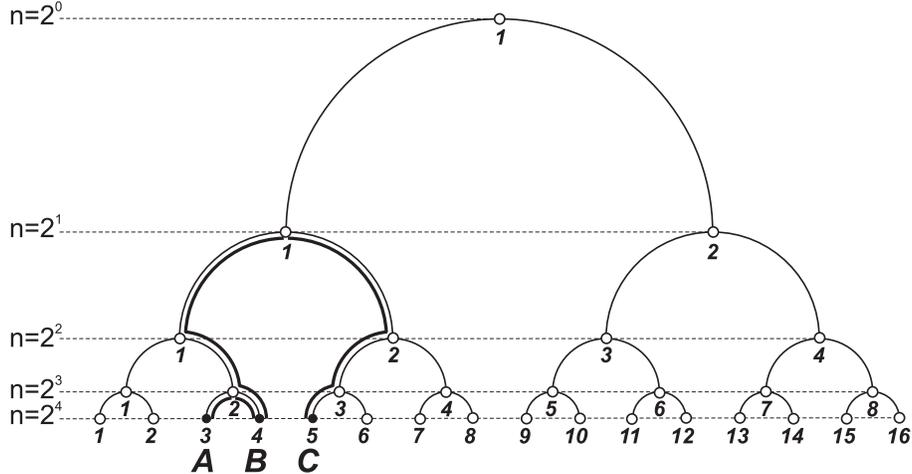,width=12cm} 
\end{center} 
\caption{The 3--branching Cayley tree. The ultrametric distance between points 
$A,B$ and $B,C$ are shown by bold lines; $n$ is the number of states in each 
generation of the tree, the maximal number of generations $L=4$.} \label{fig:3a} 
\end{figure}

The replica symmetric breaking (RSB) scheme \cite{parisi} has appeared in the 
spin--glass theory as a self--consistent approach free of many lacks of the 
replica symmetric consideration. Later on it has been realized that the 
RSB--structure naturally appears in the problem of diffusion on the boundary 
of the Cayley tree \cite{os,bachas}, where the neighboring sites are separated 
by the barriers hierarchically organized according their ultrametric distances 
on the Cayley tree. For example the neighboring points $A,B$ and $B,C$ at the 
boundary of the 3--branching Cayley tree---see fig.\ref{fig:3a}---are 
separated by the barriers depending on the ultrametric distance between points 
$A,B$ and $B,C$ on the tree.

Recall that two points, say, $B$ and $C$ in fig.\ref{fig:3a} are separated by 
the potential barrier depending on the number of the Cayley tree generations 
from the points $B$ and $C$ up to their common "parent branch". Let us label 
the states at the $L$'s generation of the 3--branching Cayley tree by the 
integer number $k$ ($1\le k\le 2^L$). If we represent $k$ by the binary 
sequence, then the ultrametric distance between points $B$ and $C$ shall 
coincide with the highest distinct rank in the binary writing of $k_B=4$ and 
$k_C=5$. 

\begin{figure}[ht] 
\begin{center} 
\epsfig{file=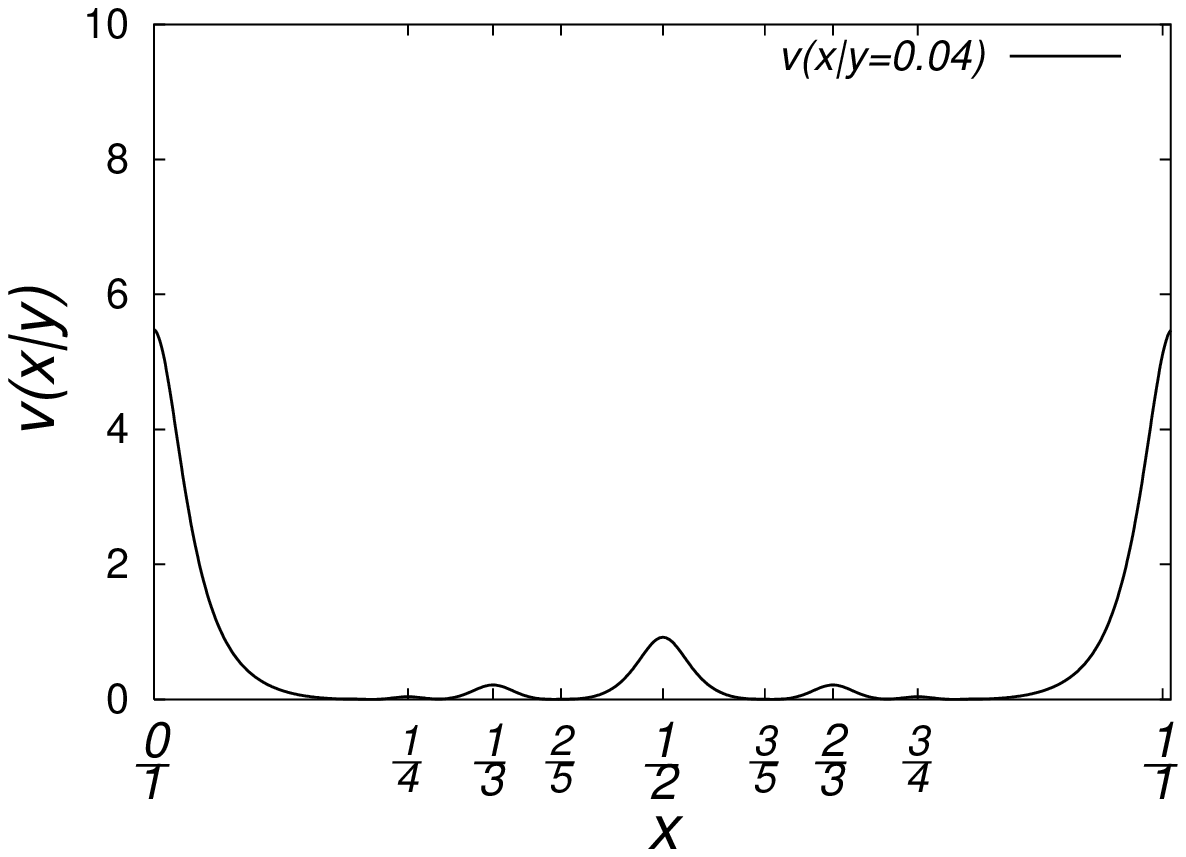,width=5.3cm} 
\epsfig{file=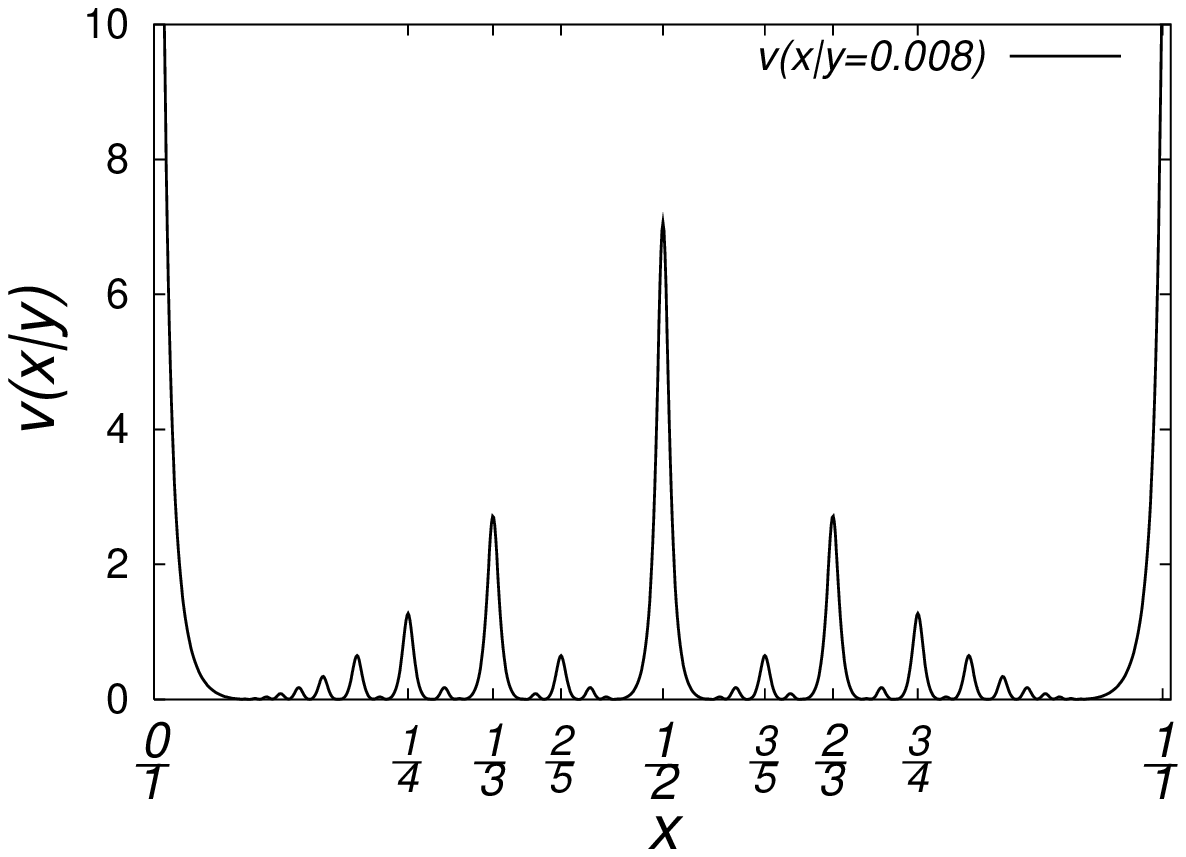,width=5.3cm} 
\epsfig{file=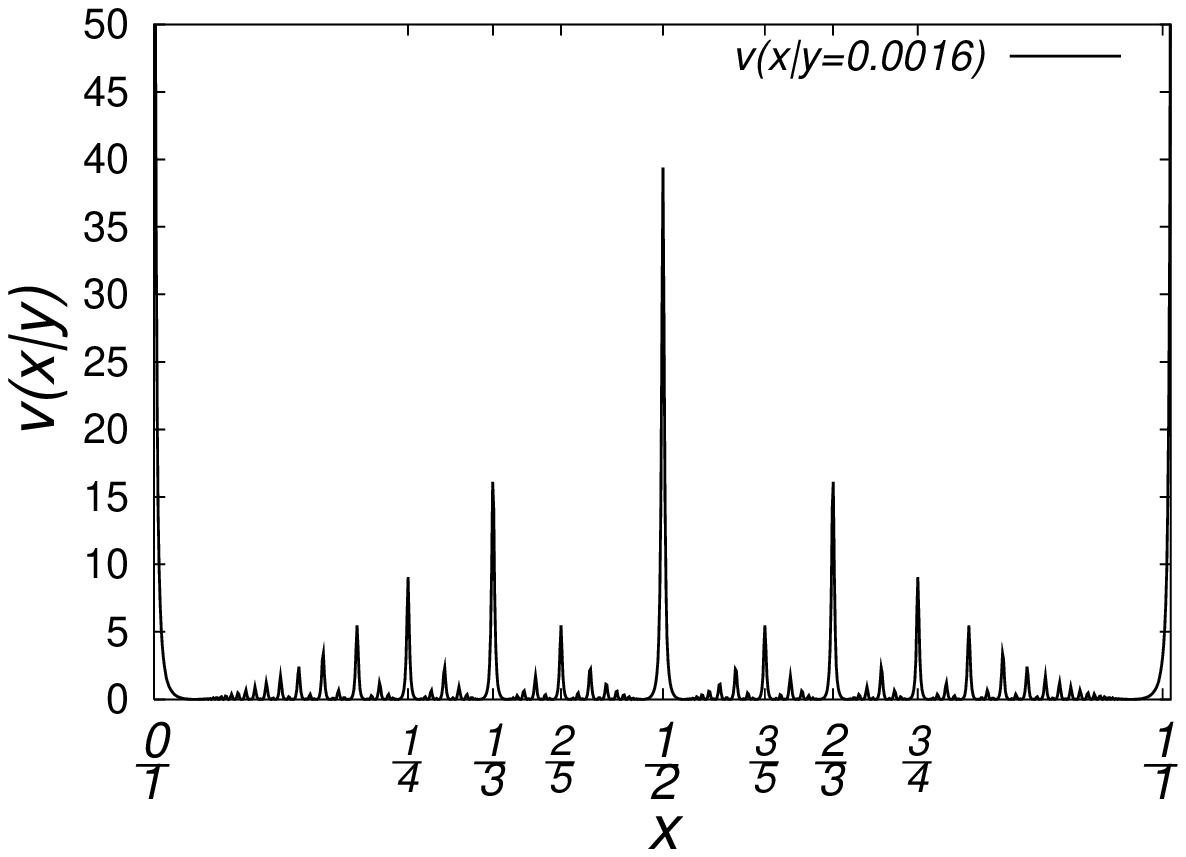,width=5.3cm} 
\end{center} 
\caption{Typical shape of the function $v(x|y)=-\ln f(x)$ for few values 
$y=0.04,\,0.008,\,0.0016$.} 
\label{fig:5} 
\end{figure}

Denote by $q_m$ the Boltzmann weights at the boundary of 3-branching Cayley 
associated to the barriers of ultrametric height $m$. All the barriers can be 
encoded in the $2^L\times 2^L$ RSB matrix ${\bf Q}(q_0,q_1,...,q_L)$ having 
the structure shown in eq.(\ref{eq:Q}) for $L=3$. Keeping in mind the 
connection to the spin glasses, the total number of states $2^L$ is associated 
to the total number of replicas, $n$, in the standard RSB scheme. 
\be 
{\bf Q}=\begin{array}{|c|c|} \hline \begin{array}{cc|cc} q_0 & q_1 & q_2 & q_2 
\\ q_1 & q_0 & q_2 & q_2 \\ \hline q_2 & q_2 & q_0 & q_1 \\ q_2 & q_2 & q_1 & 
q_0 \end{array} & \begin{array}{cccc} q_3 & q_3 & q_3 & q_3 \\ q_3 & q_3 & q_3 
& q_3 \\ q_3 & q_3 & q_3 & q_3 \\ q_3 & q_3 & q_3 & q_3 \end{array} \\ \hline 
\begin{array}{cccc} q_3 & q_3 & q_3 & q_3 \\ q_3 & q_3 & q_3 & q_3 \\ q_3 & 
q_3 & q_3 & q_3 \\ q_3 & q_3 & q_3 & q_3 \end{array} & \begin{array}{cc|cc} 
q_0 & q_1 & q_2 & q_2 \\ q_1 & q_0 & q_2 & q_2 \\ \hline q_2 & q_2 & q_0 & q_1 
\\ q_2 & q_2 & q_1 & q_0 \\ \end{array} \\ \hline 
\end{array} 
\label{eq:Q}
\ee 
The values $p_m=\frac{1}{n}q_m$ ($m=1,...,L$) define the jumping probabilities 
from some vertex $k_1$ to another vertex $k_2$  at the boundary of 3-branching 
Cayley tree separated by the barrier of ultrametric height $m$, while the 
value $p_0$ is assigned for the probabilities to stay in a given vertex. In 
each line of the matrix ${\bf Q}$ the sum of probabilities is equal to 1. In 
most physically important situations the Boltzmann weights $q_m$ depend on $m$ 
either exponentially or polynomially: 
\be 
q_m=\left\{\begin{array}{l} {\rm const}\;m^{\alpha} \\ {\rm 
const}\;e^{-\beta\, m} \end{array}\right. 
\label{eq:10} 
\ee 
Later on we pay attention only to the exponential dependence of $q_m$ on $m$, 
i.e. to the case when $q_m={\rm const}\;e^{-\beta\, m}$ (where $\beta$ stands 
for the inverse temperature). The probability $p_0$ to stay in a given vertex 
can be explicitly computed from the conservation condition $\sum_{m=0}^n 
P_m=1$ and for $q_m={\rm const}\;e^{-\beta\, m}$ it reads: 
\be 
p_0=1-\frac{1}{n}\sum_{j=1}^{L}2^j\,e^{-\beta\,j}= 
1-\frac{2(1-2^L\,e^{-\beta\,L})}{n(2-e^{\beta})} 
\label{eq:p0} 
\ee 
In order to understand the connection between the ultrametric structure of the 
barriers in case of 3--branching Cayley tree and its continuous analog 
isometrically embedded in ${\cal H}(z|{\rm Im}\,z>0)$, let us begin with the 
following observation. We can consider the function $v(x|y)=-\ln f(x|y)$ 
(where $x={\rm Re}\,z$) as the potential relief at the boundary of the Cayley 
tree cut at the distance $y={\rm Im}\,z$ from the real axis. The typical 
shapes of the function $v(x|y)$ for few fixed values $y=0.04,\,0.008,\,0.0016$ 
are shown in fig.\ref{fig:5}. This picture clearly demonstrates the 
ultrametric organization of the barriers separating the valleys. 

As one can see in fig.\ref{fig:5}, the distance from the real axis, $y$, 
characterizes the number of the Cayley tree generation: as closer is the value 
$y$ to the real axis, as more new generations of barriers appear (and as 
higher are the barriers). The relation of the system of barriers shown in 
fig.\ref{fig:5} with the structure of RSB matrix (\ref{eq:Q}) becomes now very 
straightforward. The barriers of smallest scale correspond to the values $q_1$ 
in the matrix ${\bf Q}(q_0,q_1,...,q_L)$, the barriers of the next scale 
correspond to $q_2$ etc... Hence the distance $y$ might be considered as the 
parameter controlling the size of the RSB matrix: as $y\to 0$ the size of 
${\bf Q}$ tends to infinity. Thus, one can conjecture that the number of the 
replicas, $n$, is the function of $y$. 

This relation shall be discussed in more details at length of the next 
section. The function $v(x)=-\ln f(x)$, defined on the interval $0<x<1$, has 
the properties borrowed from the structure of the underlying modular group 
$PSL(2,\mathbb{Z})$ acting in the half--plane ${\cal H}(z|y>0)$ 
\cite{magnus,beardon}. In particular: 

\begin{itemize} 
\item The local maxima of the function $v(x)$ are located at the rational points; 
\item The highest barrier on a given interval $\Delta x = [x_1,x_2]$ is 
located at a rational point $\frac{p}{q}$ with the lowest denominator $q$. On 
a given interval $\Delta x = [x_1,x_2]$ there is only one such point. The 
locations of the barriers with the consecutive heights on the interval $\Delta 
x$ are organized according to the group operation: 
\be 
\frac{p_1}{q_1}\oplus \frac{p_2}{q_2} = \frac{p_1+p_2}{q_1+q_2} 
\label{eq:rule} 
\ee 
\end{itemize} 

The figures \ref{fig:5} clarifies this statement. The highest barriers on the 
interval $0\le x\le 1$ are located at the points $x_0=0$ and $x_1=1$. 
Rewriting $0$ and $1$ correspondingly as $\frac{0}{1}$ and $\frac{1}{1}$ we 
can find the point $x_2$ of location of the barrier with the next maximal 
hight. Namely, $x_2= \frac{0}{1}\oplus 
\frac{1}{1}=\frac{0+1}{1+1}=\frac{1}{2}$. Continuing this construction we 
arrive at the hierarchical  structure of barriers located at rational points 
organized  in the Farey sequence. According to our construction this should 
replace the standard RSB scheme. The corresponding "modular" (MRSB) matrix 
shall be constructed in the next section. 

\subsection{Explicit organization of barriers on the basis of Dedekind function} 

The said above gives the idea of the construction. However to establish the 
precise connection between the standard RSB scheme and the "number--theoretic" 
MRSB scheme, we should construct the MRSB scheme with exponential organization 
of the Boltzmann weights---as in RSB. Recall that the ultrametric distance 
between two points $A$ and $B$ on the Cayley graph is equal to one half of the 
number of steps of the shortest path connecting these points. For example, the 
ultrametric distances $r_{AB}$ (between points $A$ and $B$) and $r_{AC}$ 
(between points $A$ and $C$) in fig.\ref{fig:3a} are as follows: $r_{AB}=1$ 
and $r_{AC}=3$. It is clearly seen that the points $A$ and $B$ (as well as the 
points $B$ and $C$) are separated by the barrier located at the point 
$x_1=\frac{1}{3}$, while the points $A$ and $B$ are separated by the barrier 
located at the point $x_2=\frac{2}{7}$. According to the last of 
eqs.(\ref{eq:10}), the height $U_m$ of the barrier between two points 
separated by the ultrametric distance $m$ in the standard RSB scheme reads 
\be 
U_m^{RSB} \equiv -\ln q_m=\beta\,m 
\label{eq:11} 
\ee 
The dependence of Boltzmann weights on the ultrametric distance in the MRSB 
scheme should also satisfy (\ref{eq:11}). This implies that the heights 
$U_{MRSB}$ of the MRSB--barriers should linearly depend on the ultrametric 
distance between two vertices of the Cayley tree. To adjust the heights of the 
barriers in MRSB and RSB schemes some auxiliary work has to be done. The 
corresponding construction is described in the rest of this section. Begin 
with the standard RSB tree and distribute barriers separating all $2^L$ 
boundary points of the Cayley tree in the generation $L$ equidistantly in the 
unit interval $[0,1]$. For example, for $L=1$ there are 2 vertices of the 
Cayley tree separated by a single barrier located at the point 
$x_1=\frac{1}{2}$. The barriers of the second generation $L=2$ are placed at 
the points $x_2^{(1)}=\frac{1}{4}$ and $x_2^{(2)}= \frac{3}{4}$. For $L=3$ we 
have $x_3^{(1)}=\frac{1}{8}$, $x_3^{(2)}=\frac{3}{8}$, 
$x_3^{(3)}=\frac{5}{8}$, $x_3^{(3)}=\frac{7}{8}$ etc. The points 
$x_L^{(j)}=\frac{j}{2^L}$ (where $j$ is odd and $1<j<2^L$) cover uniformly the 
unit interval $[0,1]$. Let us normalize the barrier heights as follows. In the 
finite Cayley tree of $L$ generations there are $2^{L-1}$ {\it smallest} 
barriers; all of them have equal heights. Demand these barriers be of height 
$U(L)=\frac{1}{L}$. Then the {\it averaged} barriers in some intermediate 
generation $m$ ($1\le m \le L$) have the height 
\be 
U(m,L)=\frac{L-m+1}{L}\equiv 1-\frac{m-1}{L} 
\label{eq:barrsb} 
\ee 
Hence, in the first generation $m=1$ the height of a single barrier located at 
the point $x_1=\frac{1}{2}$ is normalized by 1: $U(m=1,L)=1$. This sets our 
normalization condition. Let us note that the choice of the normalization of 
the barriers might be considered as the renormalization of the inverse 
temperature $\beta\to\tilde{\beta}L$ (see (\ref{eq:11})). The "physical" sense 
of such renormalization consists in the following. When the number of Cayley 
tree generations $L$ is increased, the corresponding landscape acquires more 
and more small--scale details without changing the height of the maximal 
barrier, which is always normalized by 1. 

Now we proceed in the similar way with the continuous MRSB--structure. Recall 
that the logarithm of the Dedekind function has maxima at the rational points 
$x=\frac{p}{q}$. The set of these points $\{x_{i}\}$ can be generated 
recursively as it has been explained above in connection with 
eq.(\ref{eq:rule}). We start with two "parent" points $0=\frac{0}{1}$ and 
$1=\frac{1}{1}$ of zero's generation. On the first step we generate a barrier 
at the point $\frac{1}{r2}=\frac{0+1}{1+1}$; on the second step we generate 
two barriers at the points $\frac{1}{3}=\frac{0+1}{1+2}$ and 
$\frac{2}{3}=\frac{1+1}{2+1}$ and so on... The number of recursive steps 
necessary to arrive at a barrier located at a specific point $x=\frac{p}{q}$ 
we shall call {\it the generation} in which this barrier has appeared for the 
first time. 

\begin{figure}[ht] 
\begin{center} 
\epsfig{file=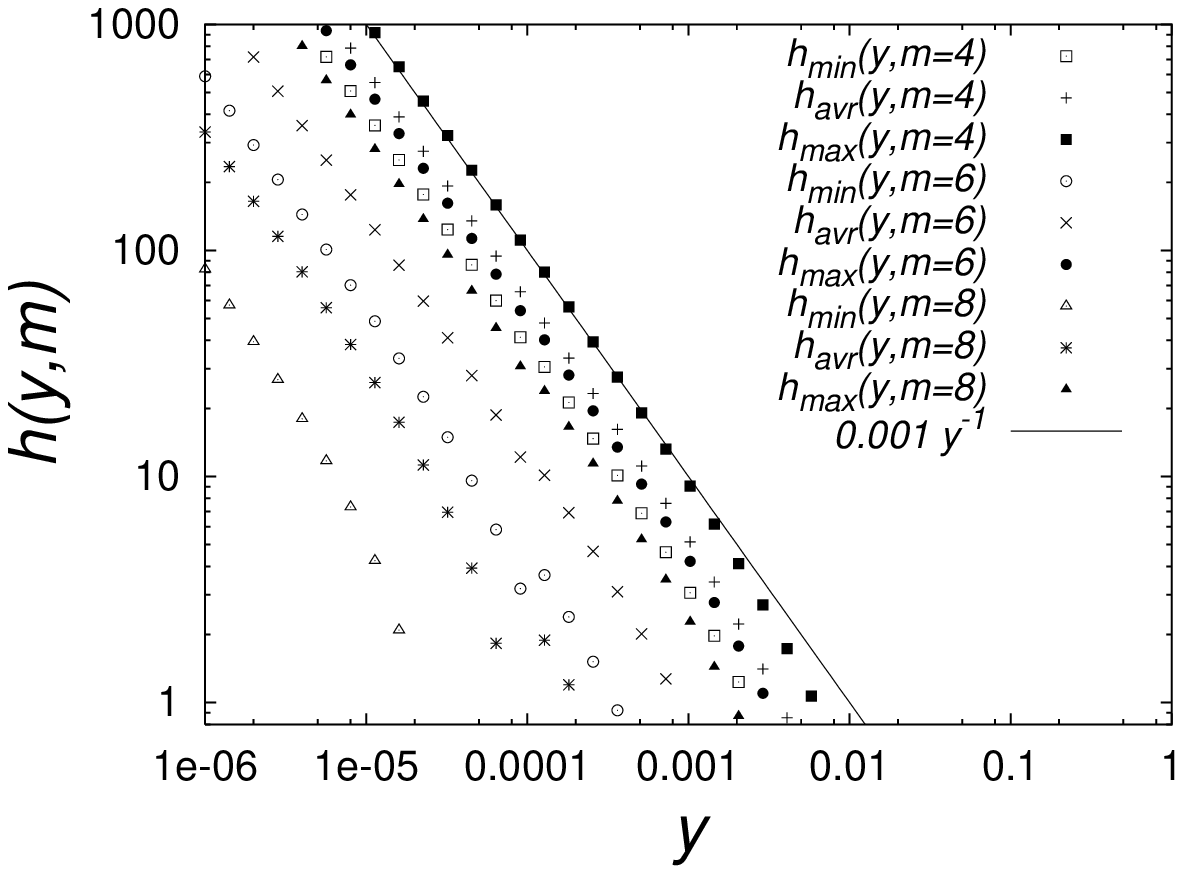,width=8cm} 
\epsfig{file=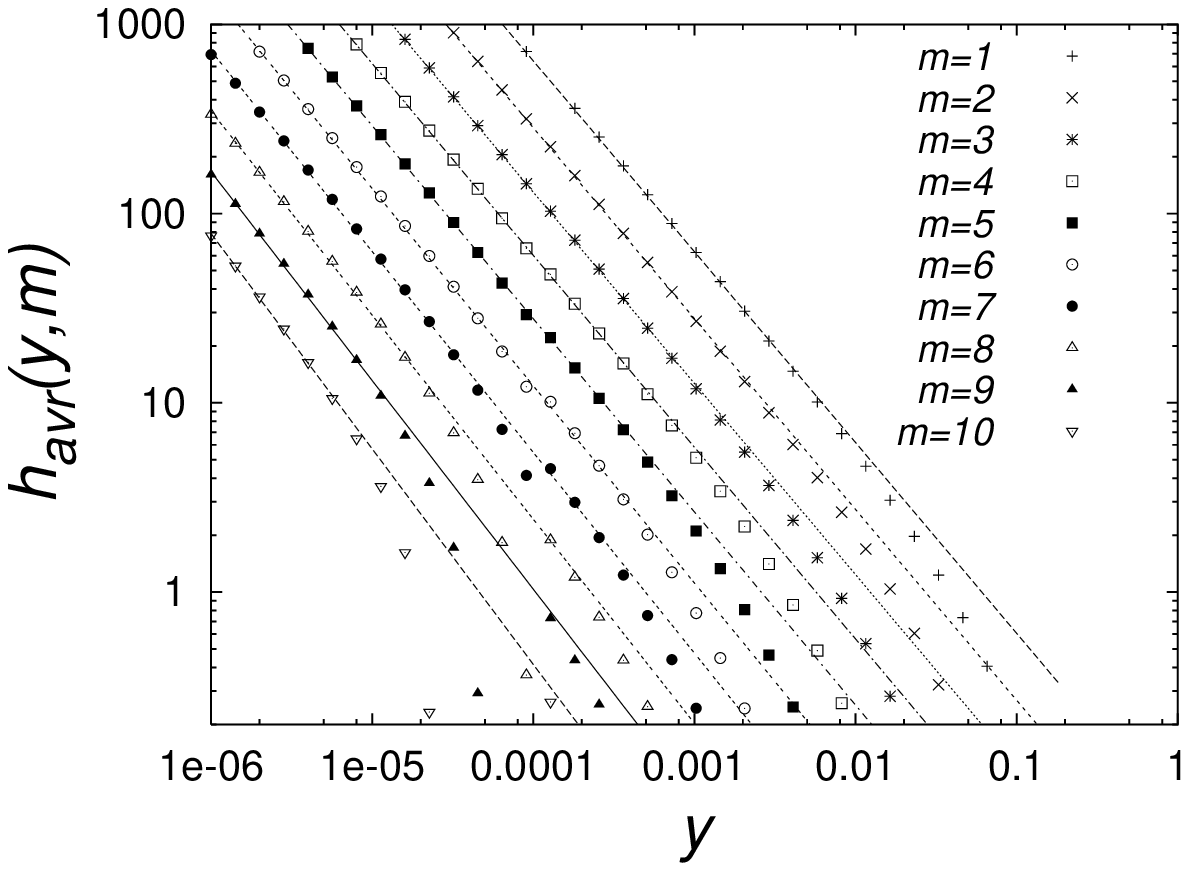,width=8cm} 
\end{center} 
\caption{a) Minimal $h_{min}(y)$, average $h_{avr}(y)$ and maximal 
$h_{max}(y)$ barriers for $m=4,6,8$; b) The function $h_{avr}(y)$ for 
$m=1,...,10$.} 
\label{fig:avr} 
\end{figure}

Let us fix the maximal generation $L$. Define $\{x_{i}\}(m)$ the set of points 
in the generation $m\le L$ at which the barriers are located. There are 
$2^{m-1}$ such points. For example, $\{x_{i}\}(m=1)= \{\frac{1}{2}\}$, 
$\{x_{i}\}(m=2)= \{\frac{1}{3}, \frac{2}{3}\}$, $\{x_{i}\}(m=3)= 
\{\frac{1}{4}, \frac{2}{5},\frac{3}{5}, \frac{3}{4}\}$ and so on. Now we need 
to express the heights $U_{MRSB}(x_{i},m,y)$ of the barriers up to the 
generation $m\le L$ via the Dedekind function $\eta(x_{i},y)$. Take the 
logarithm of the normalized Dedekind function 
\be 
h(x,y)=-\ln f(x,y) 
\label{eq:log} 
\ee 
as the possible basis for such construction (the function $f(x,y)$ is defined 
in (\ref{eq:5})). Our choice is justified by the fact that this function has 
maxima at the points $\{x_{i}\}(m)$ and demonstrates the ultrametric behavior. 
Starting from $m\ge 3$, the values of the function $h(x_{i},m,y)$ at the 
points of maxima $\{x_{i}\}(m)$ become not equal. In fig.\ref{fig:avr}a we 
have plotted the minimal $h_{min}(m,y)={\rm min}(h(x_{i},m,y))$, the average 
$h_{avr}(m,y)= \left<h(x_{i},m,y) \right>$ and the maximal $h_{max}(m,y)={\rm 
max}(h(x_{i},m,y))$ barrier heights in each of three generations $m=4,6,8$. 
The brackets $\left<\dots\right>$ denote averaging over all barriers in the 
generation $m$. The function $0.001\,y^{-1}$ is plotted for comparison in the 
same figure. 

The figure \ref{fig:avr}a allows us to conclude that $h(m,y)\sim y^{-1}$ for 
fixed $m$. More detailed information about the function $h_{avr}(m,y)= 
\left<-\ln(\eta(x_{i},y)) \right>$ can be extracted from the 
fig.\ref{fig:avr}b, which is similar to the fig.\ref{fig:avr}a but is extended 
up to $m\in[1,10]$. The data shown in fig.\ref{fig:avr}b is approximated by 
the natural fit $$ h_{avr}=a_{m}y^{b_{m}} $$ for all points such that 
$h_{avr}>10$. The results of our approximation are shown in the Table 
\ref{tab:data}a as well as are drawn in fig.\ref{fig:avr}b by lines. 

\begin{table}[ht] 
\begin{tabular}{|l|l|l|} \hline 
$~m$ & $~a_{m}$ & $~b_{m}$ \\  \hline ~1 & ~0.0594(8) & ~-1.008(2) \\ \hline 
~2 & ~0.0263(4) & ~-1.008(1) \\ \hline ~3 &~0.0118(2) & ~-1.011(1) \\ \hline 
~4 & ~0.0053(1) & ~-1.014(2) \\ \hline ~5 & ~000230(7) & ~-1.021(3) \\ \hline 
~6 & ~0.00084(7) & ~-1.04(1) \\ \hline ~7 & ~0.00031(4) & ~-1.06(1) \\ \hline 
8 & ~0.000118(17) & ~-1.08(6) \\ \hline ~9 & ~0000040(7) & ~-1.102(1) \\ 
\hline ~10 & ~0000012(3) & ~-1.14(2) \\ \hline 
\end{tabular} \hspace{2in} 
\vspace{-2.73in} 

\begin{tabular}{|l|l|} \hline  
$~y^{*}$  & $~c(y^{*})$ \\ \hline 
$~5.12 \times 10^{-5}$ & ~0.820(11) \\ \hline $~2.56 \times 10^{-5}$ & 
~0.8057(99) \\ \hline $~1.28 \times 10^{-5}$ & ~0.7944(86) \\ \hline $~6.4 
\times 10^{-6}$ & ~0.7718(29) \\ \hline $~3.2 \times 10^{-6}$ & ~0.7623(22) \\ 
\hline $~1.6 \times 10^{-6}$ & ~0.7579(27) \\ \hline $~8 \times 10^{-7}$ & 
~0.7556(30) \\ \hline $~4 \times 10^{-7}$ & ~0.7544(32) \\ \hline $~2 \times 
10^{-7}$ & ~0.7538(37) \\ \hline $~1\times 10^{-7}$ & ~0.7535(34) \\ \hline 
\end{tabular} 
\hspace{-2.5in} 
\caption{a) Numerical results of the fit $h_{avr}=a_{m}y^{b_{m}}$; b) 
Dependence of the parameter $c(y)$ on $y=y^{*}$ up to $L=10$ generations---see 
eq.\ref{eq:defbar}.} 
\label{tab:data} 
\end{table} 

\begin{figure}[ht] 
\begin{center} 
\epsfig{file=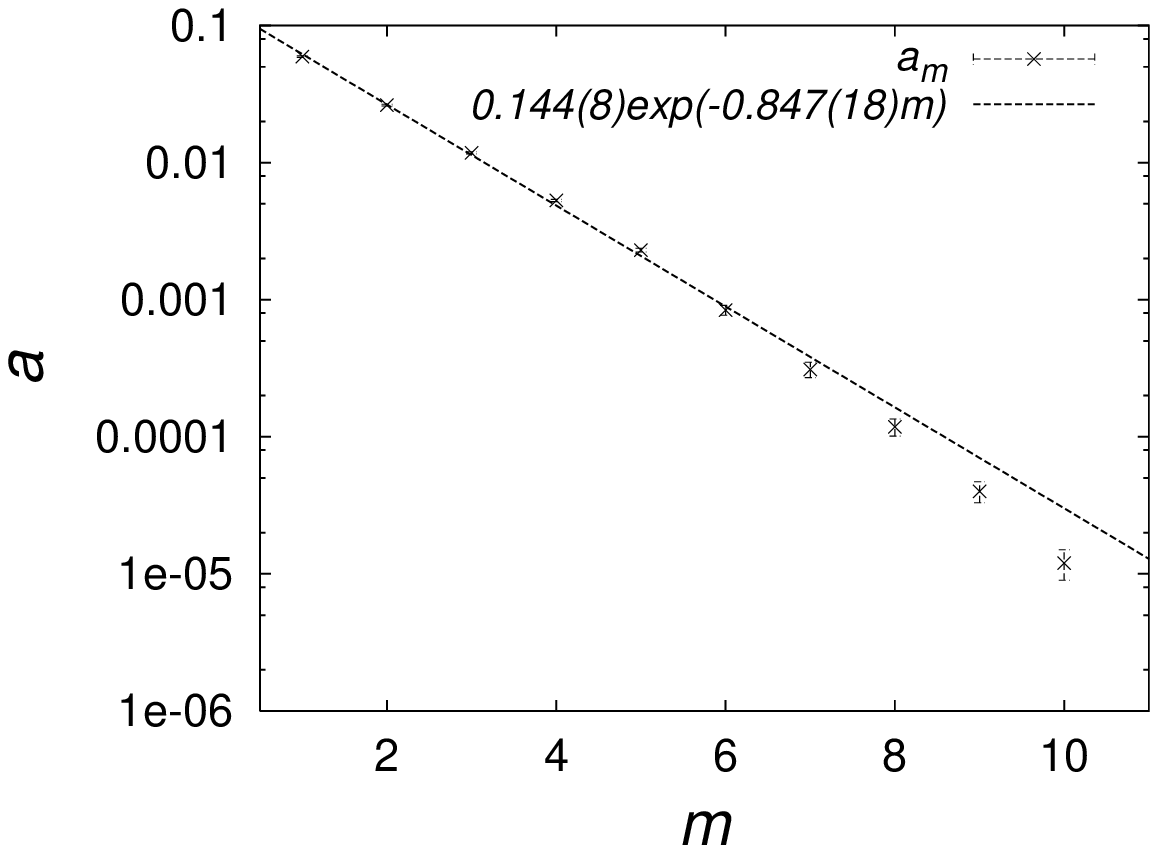, width=8cm} 
\epsfig{file=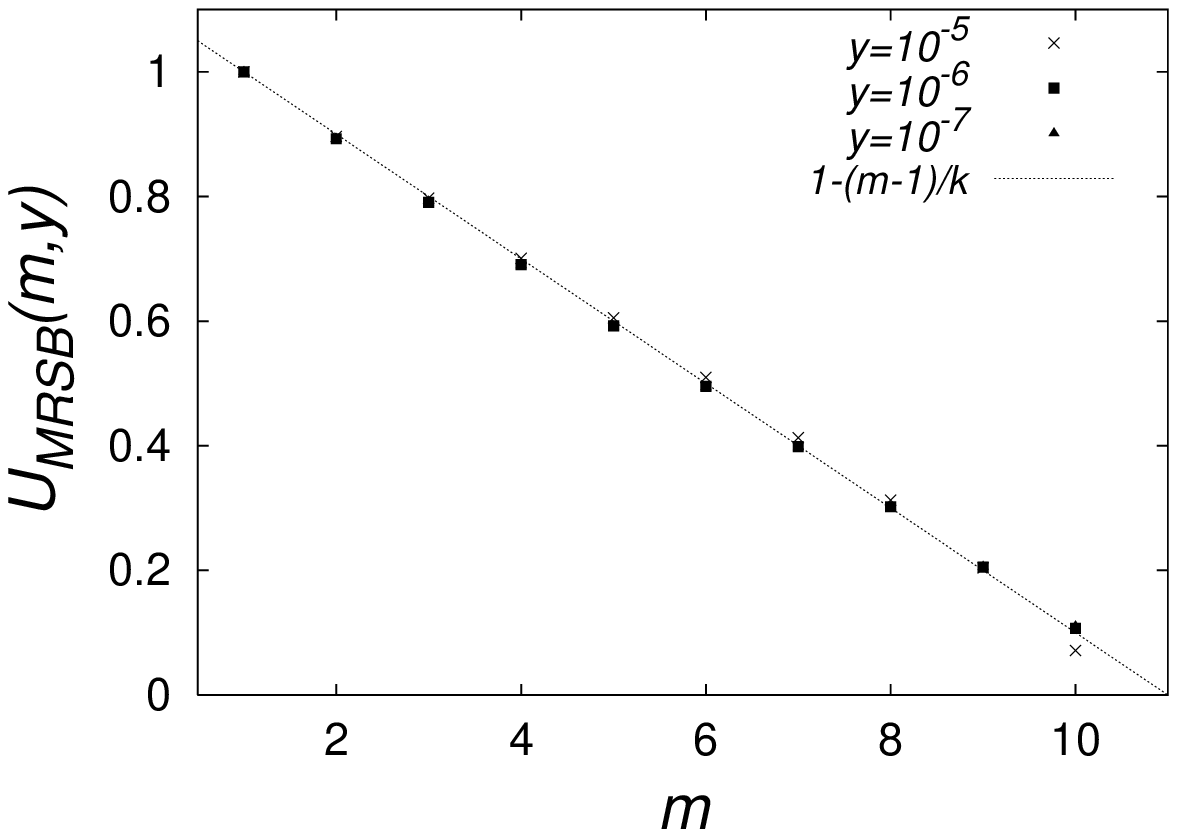,width=8cm} 
\end{center} 
\caption{a) Approximation of the coefficient $a_{m}$; b) Ultrametric barriers 
$U_{MRSB}(m,y)=\frac{1}{c(y)\,L}\ln \frac{h(m,y))}{h(1,y)}$.} 
\label{fig:data} 
\end{figure}

We have plotted the coefficient $a_{m}$ as a function of $m$ in 
fig.\ref{fig:data}a. The approximation $a_{m}\simeq 
0.144(8)\,e^{-0.847(18)\,m}$ is shown in the same figure by dashed line. 
Hence, we expect the following relation $h(m,y) \simeq 0.144(8)\, 
e^{-0.847(18)\,m}\, y^{-1}$. 

The analysis of the numerical data for the averaged logarithm of the Dedekind 
function allows us to conjecture the general ansatz 
\be 
h_{avr}(m,y)\equiv \left<-\ln f(x_{i},y) \right> = d(y)\,e^{-c(y)\,m}\, y^{-1} 
\label{eq:avrapp} 
\ee 
This equation establishes the functional dependence of the average height 
$h_{avr}$ on the generation $m$ ($m=1,2,...$) for fixed $y$. Recall that the 
generation $m$ is defined as the minimal number of recursive steps necessary 
to arrive at a barrier located at a specific point $x_i=\frac{p}{q}$. Let us 
fix now the maximal generation $L$. Our desire is to obtain the normalized 
barrier for MRSB scheme like it has been done for RSB (see 
eq.(\ref{eq:barrsb})). Taking into account that (\ref{eq:avrapp}) is valid for 
any $m\ge 1$, we may extract $m$ from (\ref{eq:avrapp}) and write it as 
follows 
\be 
m=1+\frac{1}{c(y)}\ln \frac{h_{avr}(m,y)}{h_{avr}(m=1,y)} 
\label{eq:m} 
\ee 
Substituting (\ref{eq:m}) into (\ref{eq:barrsb}) we get the following formal 
expression for the height of the barriers in the MRSB scheme 
\be 
U_{MRSB}(x_{m},y,L)\equiv 1-\frac{m-1}{L}=1+\frac{1}{c(y)\,L} \ln 
\frac{h_{avr}(m,y)}{h_{avr}(1,y)} 
\label{eq:umrsb} 
\ee 
Now we should explain how it is possible to use the formal equation 
(\ref{eq:umrsb}) in practical computations. Let us fix the maximal generation 
$L$ and define the value $y^*$ from the equation 
\be 
h_{avr}(y^*,L)=1 
\label{eq:y_star} 
\ee 
The value $y^*$ has the following sense. For all $y<y^*$ the value 
$h_{avr}(m,y)$ does not depend on $m$ ($m\le L$) because all barriers in the 
generation $m$ are already presented and hence the average hight $h_{avr}$ can 
not be changed for any $y<y^*$. Therefore we can postulate, that for all 
$y<y^*$ the height of the barrier is described by the function 
\be 
U_{MRSB}(x_{i})=1+\frac{1}{c(y^*)\,L}\ln \frac{\ln f\left(x_{i}, y^*\right)} 
{\ln f\left(\frac{1}{2}, y^*\right)} 
\label{eq:defbar} 
\ee 
where $x_i$ is the rational point of the barrier location and we have used 
(\ref{eq:log}). Special attention deserves the discussion of the dependence 
$c(y^*)$. As it is shown in the Table \ref{tab:data}b, for sufficiently small 
$y^*$ the value of $c$ tends to the constant value. Thus the function 
$U_{MRSB}(x_{i})$ defines the heights of ultrametric barriers in MRSB scheme 
fully consistent with the heights of the RSB--barriers for the Cayley tree. It 
should be noted that for intermediate values of $y$ some barriers of large 
generations are negative due to the cut (\ref{eq:y_star}), however having $m$, 
we can always find $y^*$ such that all barriers are positive. For example, for 
$y \le y^*=10^{-6}$ all barriers up to $m=10$ generations are positive. 

\begin{figure}[ht] 
\begin{center} 
\epsfig{file=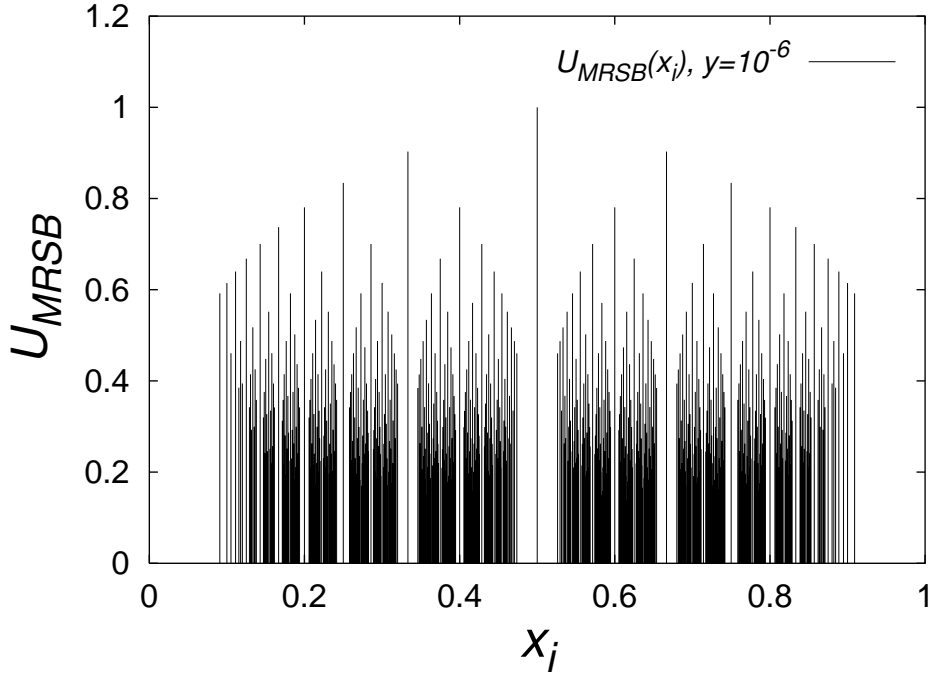,width=13cm} 
\end{center} 
\caption{Ultrametric barriers $U_{MRSB}(x_{i})$ at the rational points $x_{i}$ 
for $y=10^{-6}$} 
\label{fig:barvid} 
\end{figure} 

Our tedious but simple construction is clearly explained in the figure 
\ref{fig:barvid}, where we have plotted the heights of barriers 
$U_{MRSB}(x_{i})$ at the rational points $x_{i}$ up to the generation $L=10$ 
for $y^*=10^{-6}$. Once the maximal generation $L$ is fixed and all barriers 
for all generations (up to the maximal one) are shown in the picture, then 
this picture is unchanged for all $y$ such that $y<y^*$. According to our 
definition, these barriers are normalized in such a way that the height of the 
largest barrier is equal to $1$. The averaged height of the barriers of second 
(third, forth, etc...) generation is equal to $0.9$ ($0.8$, $0.7$, etc...) -- 
compare to (\ref{eq:barrsb}). Let us repeat that in fig.\ref{fig:barvid} we 
have displayed the barriers corresponding to generations $m$ up to $m=L$. We 
can easily relax this condition and work with the set of {\it all} possible 
barriers up to some fixed minimal value $y_{min}$ without paying attention how 
many generations are counted. 

\subsection{Description of dynamical models} 

Our aim is to investigate numerically the probability distribution of a random 
walk on the boundary of an ultrametric space. This problem has been a subject 
of the original work \cite{os} where the dynamics in the ultrametric space was 
considered in the frameworks of the discrete model of jumps on the boundary of 
3--branching Cayley tree. The first application of $p$--adic analysis for the 
computation of the probability distribution of the diffusion at the boundary 
of the ultrametric tree has been successfully realized in \cite{avet1}. 

Below we reconsider the diffusion problem in the continuous metric space, 
where the ultrametric organization of the barriers is due to the modular 
structure of the Dedekind function. The advantages of such consideration we 
shall discuss in the Conclusion. Let us specify the models under 
consideration. 

\subsubsection{I. The standard RSB model} 

In this model we consider the random jumps at the boundary of 3--branching 
Cayley tree---see fig.\ref{fig:3a}. The transition probabilities between the 
Cayley tree vertices are encoded in the matrix (\ref{eq:Q}). As explained 
above, we fix the total number of generations $L$ and set the highest barrier 
(in the first generation) always equal to 1. Then the heights of the barriers 
of generation $m$ are given by (\ref{eq:barrsb}), while the height of the 
smallest barriers in the generation $L$ is equal to $\frac{1}{L}$. Increasing 
the number of generations $L$, we increase the resolution of our model without 
changing the value of the maximal barrier. We use this problem of diffusion on 
the boundary of the standard 3--branching Cayley tree as a testing area for 
diffusion in the continuous tree--like ultrametric space. 

Let $P_N(k,k_0)$ be the partition function of the random walk starting at the 
point $k_0$ and ending after $N$ steps at the point $k$, where $k$ labels the 
vertices of the $k$'s generation of the Cayley tree and both points $k_0$ and 
$k$ belong to the $m$'s generation of the tree. The corresponding diffusion is 
governed by the recursion relation 
\be 
\left\{\begin{array}{l} \disp P^{RSB}_{N+1}(k)=\sum_{k'=0}^{2^L-1} 
Q^{RSB}(k,k') P^{RSB}_N(k') \\ \disp P^{RSB}_{N=0}(k)=\delta_{k,k_0} 
\end{array}\right. 
\label{eq:14} 
\ee 
where $Q^{RSB}(k,k')$ is the $2^L\times 2^L$ matrix whose elements are the 
Boltzmann weights associated with jumps from some point $k_1$ to the point 
$k_2$ ($\{k_1,k_2\}\in [1,2^L]$) at the boundary of 3--branching Cayley tree 
cut at the generation $L$. In order to take into account the ultrametricity of 
the Cayley tree, the matrix $Q^{RSB}(k,k')$ should have the RSB structure and 
hence should coincide with the matrix ${\bf Q}$ in (\ref{eq:Q}). 

\subsubsection{II. The quasi-continuous MRSB model} 

\begin{figure}[ht] 
\begin{center} 
\epsfig{file=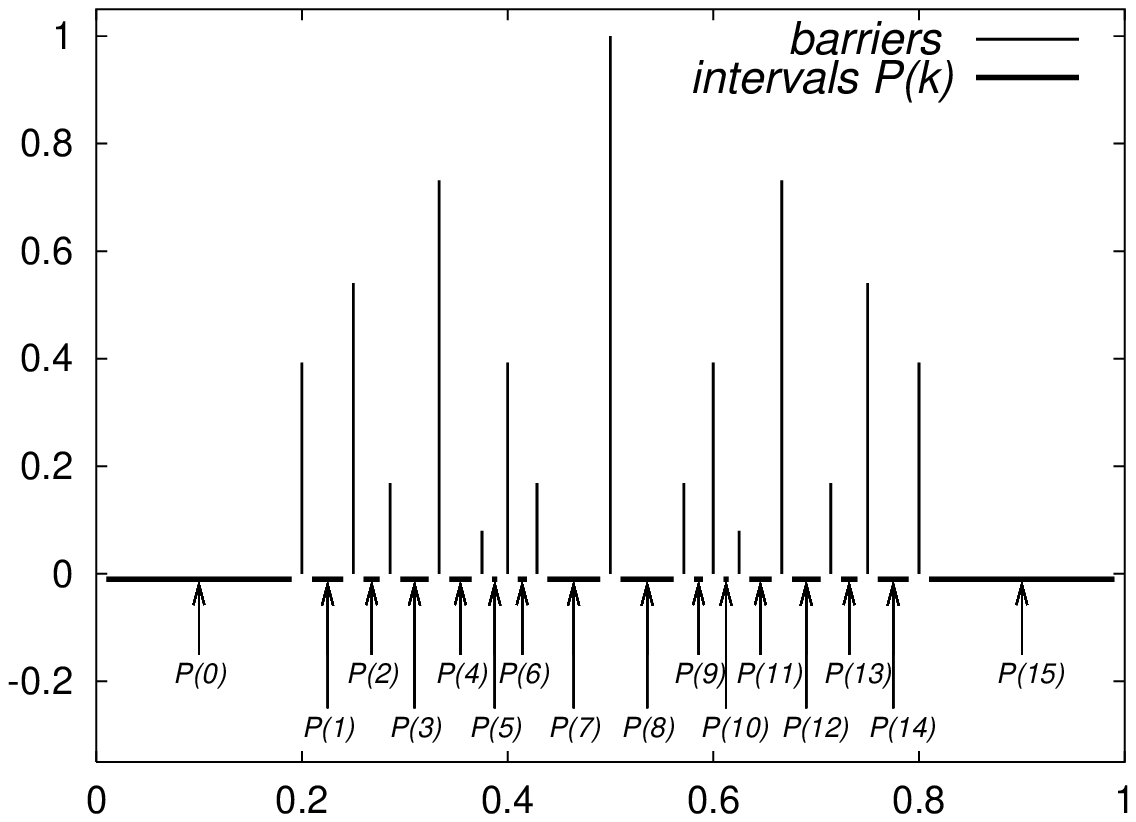,width=8cm} 
\epsfig{file=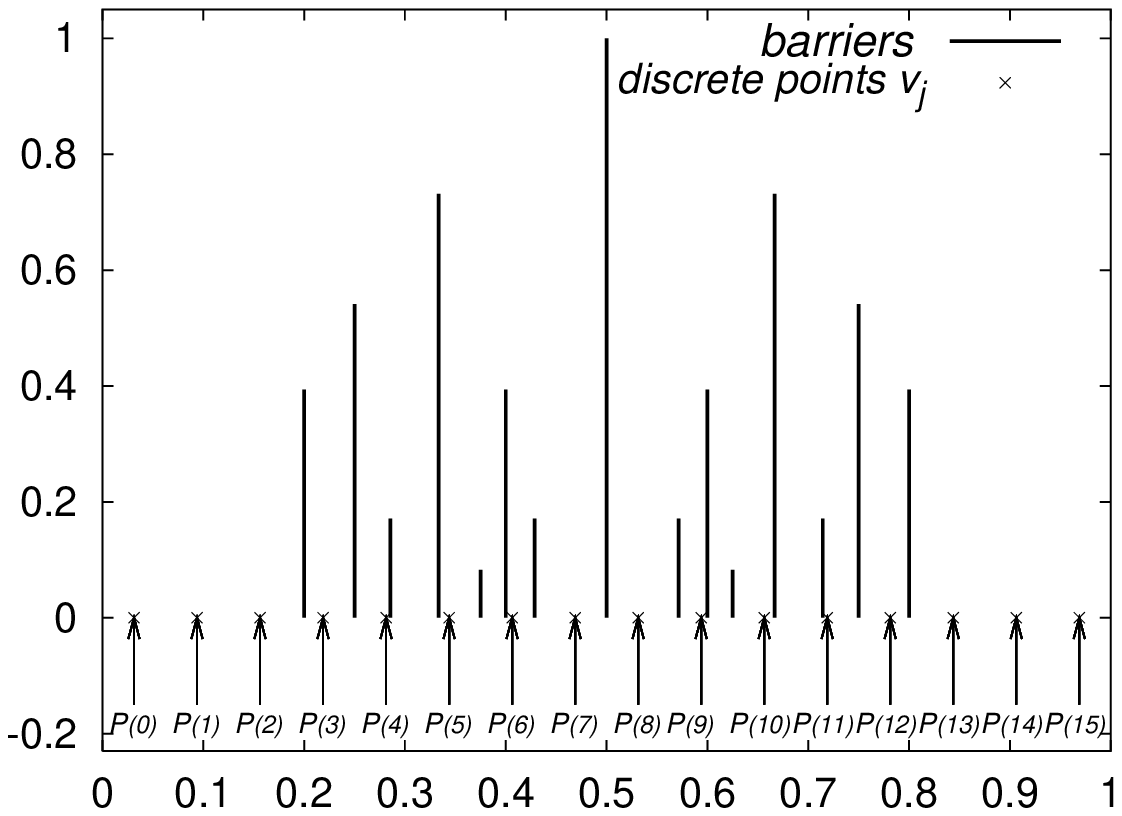,width=8cm} 
\end{center} 
\caption{a) Location of barriers up to 4 generations: a) Model I: the values 
$v_i$ correspond to the intervals between maxima; b) Model II: the values 
$v_i$ correspond to the discrete points.} 
\label{fig:barl} 
\end{figure}

This model is defined as follows. We fix the maximal generation $L$ and 
consider the heights of the potential $U^{MRSB}$ (see (\ref{eq:defbar})) 
corresponding to generations less or equal $L$. For example, in 
fig.\ref{fig:barl}a we have plotted $15=2^{L}-1=n$ barriers up to $L=4$ 
generation. The rational points $x_{i}$ split the interval $[0,1]$ into 16 
subintervals $[x_{i-1},x_{i}],\; i=1,2,...,15$ (including the subintervals 
$[0,x_{1}]$ and $[x_{15},1]$). The random walker can jump form one interval to 
any other. Let us enumerate the intervals by the variable $k=1,2,3,...,16$. 
So, $k=1$ denotes the interval $[0,x_{1}]$, $k=2$ is the interval 
$[x_{1},x_{2}]$, etc. Let $U_{MRSB}(k,k')$ be the maximal barrier between the 
intervals $k$ and $k'$ defined by (\ref{eq:defbar}). By construction 
$U^{MRSB}(k,k)=0$. According to our rules if $x \in [x_{k-1},x_{k}]$ and $x' 
\in [x_{k'-1},x_{k'}]$ then $U^{MRSB}(x,x')= U^{MRSB}(k,k')$. Thus we can 
write the recursive relation for the continuous distribution function 
$W_{N}(x)$. This function is defined for $x\in {\mathbb R},\;(0\le x\le 1)$, 
where $N$ enumerates the time moments. We claim for this quasi-continuous MRSB 
case the following form of the master equation which should replace 
eq.(\ref{eq:14}). 
\be 
\left\{\begin{array}{l} \disp W_{N+1}(x)=\int_0^1 dx'\, e^{\disp -\beta 
U_{MRSB}(x,x')} \; W_N(x') \\ \disp W_{N=0}(x)=\delta_{x,x_0} 
\end{array}\right. 
\label{eq:15} 
\ee 

We can transform (\ref{eq:15}) making it more similar to (\ref{eq:14}). 
Namely, define $P^{MRSB}_{N}(k)$ -- the probability that the walker belongs to 
the $k$-th interval on the $N$-th jump (time step) 
\be 
P^{MRSB}_{N}(k)= \int_{x_{k-1}}^{x_{k}}W_{N}(x) dx 
\label{eq:defpn} 
\ee 
The normalization of the probability defines the sum over all intervals 
$\sum_{k=1}^{n} P^{MRSB}_{N}(k) = \int_{0}^{1} W_{N}(x) dx=1$. Integrating 
(\ref{eq:15}) over the interval $[0,1]$ and replacing the integrals over $dx$ 
and $dx'$ by sums over $k$ and $k'$, we get 
\be 
\begin{array}{lll} \disp \sum_{k=1}^{n}P^{MRSB}_{N+1}(k) & = & \disp 
\int_{0}^{1} W_{N+1}(x) dx= \int_{0}^{1} \int_0^1 \, e^{\disp -\beta 
U_{MRSB}(x,x')} \; W_N(x')dx' dx \\ & = & \disp \sum_{k=1}^{n} \sum_{k'=1}^{n} 
\int_{x_{k-1}}^{x_{k}} \int_{x'_{k'-1}}^{x'_{k'}} e^{\disp -\beta 
U_{MRSB}(x,x')} W^{MRSB}_{N}(k') dx' dx \\ & = & \disp \sum_{k=1}^{n} 
\sum_{k'=1}^{n} l_{k} e^{\disp -\beta U_{MRSB}(k,k')} P^{MRSB}_{k'} 
\end{array} 
\label{eq:16n} 
\ee 
where $l_{k}$ is the length of the $k$-th interval 
$l_{k}=\int_{x_{k-1}}^{x_{k}}dx= x_{k}-x_{k-1}$. Here we have used 
(\ref{eq:defpn}) to express $P^{MRSB}_{N+1}(k)$ and $P^{MRSB}_{N}(k')$ via 
$W^{MRSB}_{N+1}(x)$ and $W^{MRSB}_{N}(x')$. Now we can separate equations for 
different $k$ and write 
\be 
P^{MRSB}_{N+1}(k) = \sum \limits_{k'=1}^{n} Q(k,k')P^{MRSB}_{N}(k') 
\ee 
where 
\be 
Q(k,k')=l_{k} e^{\disp -\beta U_{MRSB}(k,k')} 
\label{eq:qu}
\ee 
is the probability to jump from the interval $k'$ to the interval $k$. This 
probability is the product of the measure of the appropriate interval $l_{k}$ 
and the Boltzmann weight $e^{\disp -\beta U^{MRSB}(k,k')}$. Define a vector 
${\bf P}^{MRSB}_{N}$ whose elements $P^{MRSB}_{N}(k)$ are the probabilities 
for a walker to reach the interval $k$ on the $N$'s jump. The vector ${\bf 
P}^{MRSB}_{0}$ is the initial probability distribution on the intervals. 
Introduce a matrix ${\bf Q}$ with the elements $Q(k,k')$, and a matrix ${\bf 
U}^{MRSB}$ with the elements $U^{MRSB}(k,k')$. The elements of these two 
matrices are connected by the relation (\ref{eq:qu}). The probability for the 
walker to reach the interval $k$ at the time moment $N$ is the $k$'s element 
of the vector ${\bf P}^{MRSB}_{N}(k)$ 
\be 
{\bf P}^{MRSB}_{N}(k)={\bf Q}^{N}{\bf P}^{MRSB}_{0} 
\ee 
Let us remind, that the barrier $U^{MRSB}(k,k')$ between the points $k$ and 
$k'$ is the highest barrier on the interval $[k,k']$. Such barriers are 
located at the rational points $\{k_{m}\}$. This defines the ultrametric 
structure of the matrix ${\bf U}^{MRSB}$ and, therefore, of ${\bf Q}$. By 
construction the matrix ${\bf U}^{MRSB}$ reflects the ultrametric hierarchy of 
barriers appearing in the standard RSB--matrix (see eq.(\ref{eq:Q})). For 
example, the structure of the "modular" MRSB matrix ${\bf U}^{MRSB}$ defining 
the heights of the barriers between all the intervals along the $x$--axis for 
$L=8$ is displayed below: 

\renewcommand{\arraystretch}{2.6} 

\be 
{\bf U}_{MRSB}=\begin{array}{|c|c|} \hline \begin{array}{cc|cc} U_0 & 
U\left(\frac{1}{4}\right) & U\left(\frac{1}{3}\right) & 
U\left(\frac{1}{3}\right) \\ U\left(\frac{1}{4}\right) & U_0 & 
U\left(\frac{1}{3}\right) & U\left(\frac{1}{3}\right) \\ \hline 
U\left(\frac{1}{3}\right) & U\left(\frac{1}{3}\right) & U_0 & 
U\left(\frac{2}{5}\right) \\ U\left(\frac{1}{3}\right) & 
U\left(\frac{1}{3}\right) & U\left(\frac{2}{5}\right) & U_0 \end{array} & 
\begin{array}{cccc} U\left(\frac{1}{2}\right) & U\left(\frac{1}{2}\right) & 
U\left(\frac{1}{2}\right) & U\left(\frac{1}{2}\right) \\ 
U\left(\frac{1}{2}\right) & U\left(\frac{1}{2}\right) & 
U\left(\frac{1}{2}\right) & U\left(\frac{1}{2}\right) \\ 
U\left(\frac{1}{2}\right) & U\left(\frac{1}{2}\right) & 
U\left(\frac{1}{2}\right) & U\left(\frac{1}{2}\right) \\ 
U\left(\frac{1}{2}\right) & U\left(\frac{1}{2}\right) & 
U\left(\frac{1}{2}\right) & U\left(\frac{1}{2}\right) \end{array} \\ \hline 
\begin{array}{cccc} U\left(\frac{1}{2}\right) & U\left(\frac{1}{2}\right) & 
U\left(\frac{1}{2}\right) & U\left(\frac{1}{2}\right) \\ 
U\left(\frac{1}{2}\right) & U\left(\frac{1}{2}\right) & 
U\left(\frac{1}{2}\right) & U\left(\frac{1}{2}\right) \\ 
U\left(\frac{1}{2}\right) & U\left(\frac{1}{2}\right) & 
U\left(\frac{1}{2}\right) & U\left(\frac{1}{2}\right) \\ 
U\left(\frac{1}{2}\right) & U\left(\frac{1}{2}\right) & 
U\left(\frac{1}{2}\right) & U\left(\frac{1}{2}\right) \end{array} & 
\begin{array}{cc|cc} U_0 & U\left(\frac{3}{5}\right) & 
U\left(\frac{2}{3}\right) & U\left(\frac{2}{3}\right) \\ 
U\left(\frac{3}{5}\right) & U_0 & U\left(\frac{2}{3}\right) & 
U\left(\frac{2}{3}\right) \\ \hline U\left(\frac{2}{3}\right) & 
U\left(\frac{2}{3}\right) & U_0 & U\left(\frac{3}{4}\right) \\ 
U\left(\frac{2}{3}\right) & U\left(\frac{2}{3}\right) & 
U\left(\frac{3}{4}\right) & U_0 \end{array} \\ \hline \end{array} 
\label{eq:U} 
\ee 

\renewcommand{\arraystretch}{1.2} 

\subsubsection{III. The discretized MRSB model} 

In this model the location of the barriers is the same as in the model II, but 
now we split the total interval $[0,1]$ into $n=2^{L}$ subintervals of the 
equal length $\frac{1}{n}$ each. The midpoint of the $k$-th interval is the 
point $\bar x_{k}=-\frac{1}{2n}+\frac{k}{n}$. We consider the distance between 
intervals $k$ and $k'$ as the distance between the midpoints: $U_{III}(k,k')= 
U^{MRSB}(\bar x_{k},\bar x_{k'})$. Namely, we consider jumps between $n$ 
midpoints $k,\; (k=1,2,..., n)$ with coordinates $\bar 
x_{k}=-\frac{1}{2n}+\frac{k}{n}$. The probability to jump from a point $k'$ to 
a point $k$ is defined by the relation 
\be 
Q^{III}(k,k')=\frac{1}{L} e^{-\beta U^{MRSB}(\tilde x_{k},\tilde x_{k'})} 
\label{eq:II} 
\ee 
Each element $P^{III}_{N}(k)$ of the of the probability vector ${\bf P}$ 
corresponds to a point. In this model all barriers $x_{i}=\frac{p_{i}}{q_{i}}$ 
lie between the midpoints---see fig.\ref{fig:barl}b. 

Comparing the models II and III it is easy to understand, that the number $n$ 
of points $\bar x_{k}$ in the interval between barriers $[x_k,x_{k+1}]$ plays 
role of measure of that interval. 

Below we compare the numerical solutions of the models I, II and III. To be 
more specific, we consider simultaneously the $N$--step random walk on the 
Cayley tree (the model I) and on its continuous analog (the models II and 
III). First of all, we calculate the elements $[{\bf Q}^{N}](i,j)$ of the 
matrix ${\bf Q}^{N}$ for all three models. Then we compute: 

\begin{itemize} 
\item[--] The probability $P_{0}$ to stay at the initial point after $N$ 
random jumps 
\be 
P_{0}=\sum_{i=1}  [{\bf Q}^{N}](i,i) 
\label{eq:defp0} 
\ee 
\item[--] The average ultrametric distance $U$ between the initial and the 
final points of the $N$--step random walk: 
\be 
U=\sum_{i=1}^{n} \sum_{j=1}^{n} [{\bf U}](i,j) [{\bf Q}^{N}](i,j) 
\label{eq:defu} 
\ee 
where $[{\bf U}](i,j)$ is the element of the matrix ${\bf U}$ defining the 
ultrametric distances between all available points. The size of the matrices 
${\bf U}$ and ${\bf Q}$ are $n=2^{L}$, where $L$ is the total number of 
generations. 
\end{itemize} 

\subsection{Numerical results for the models I, II, III}

\begin{figure}[ht] 
\begin{center} 
\epsfig{file=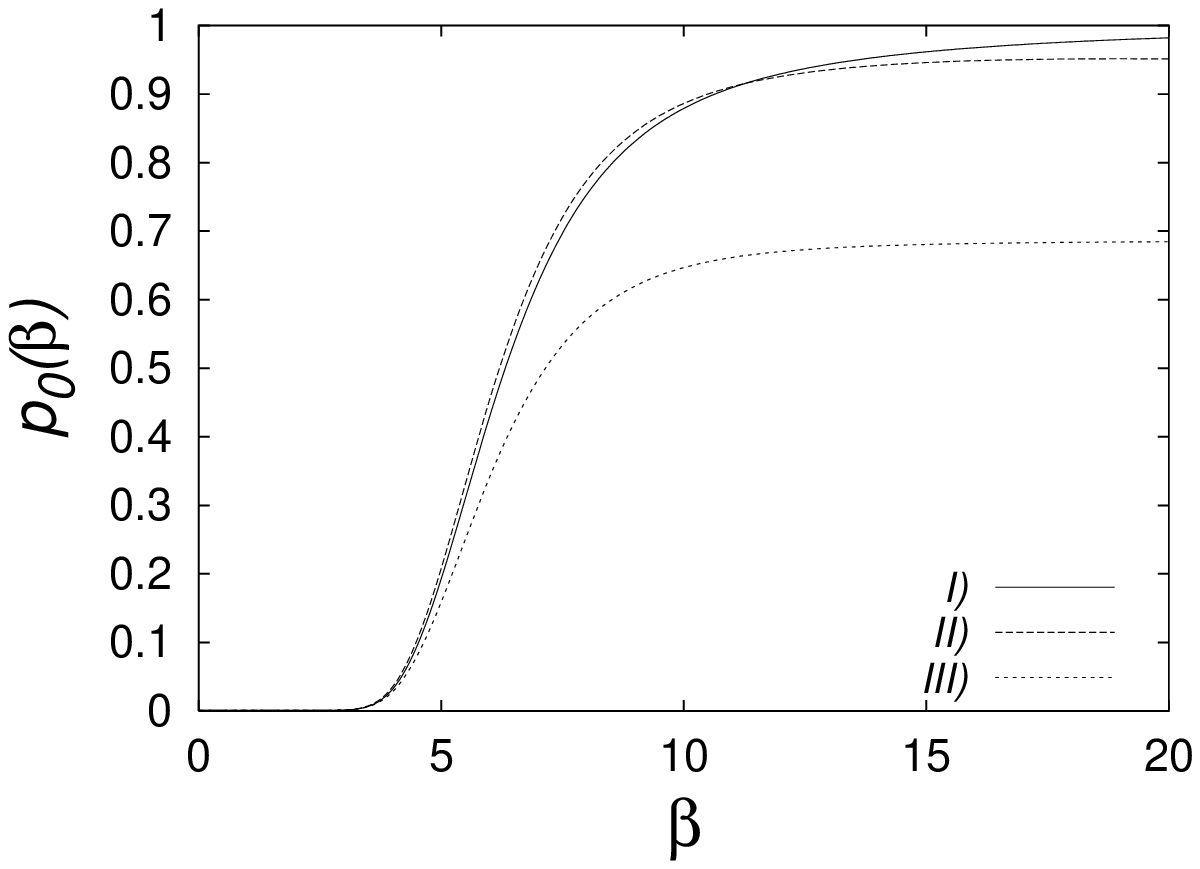,width=8cm} 
\epsfig{file=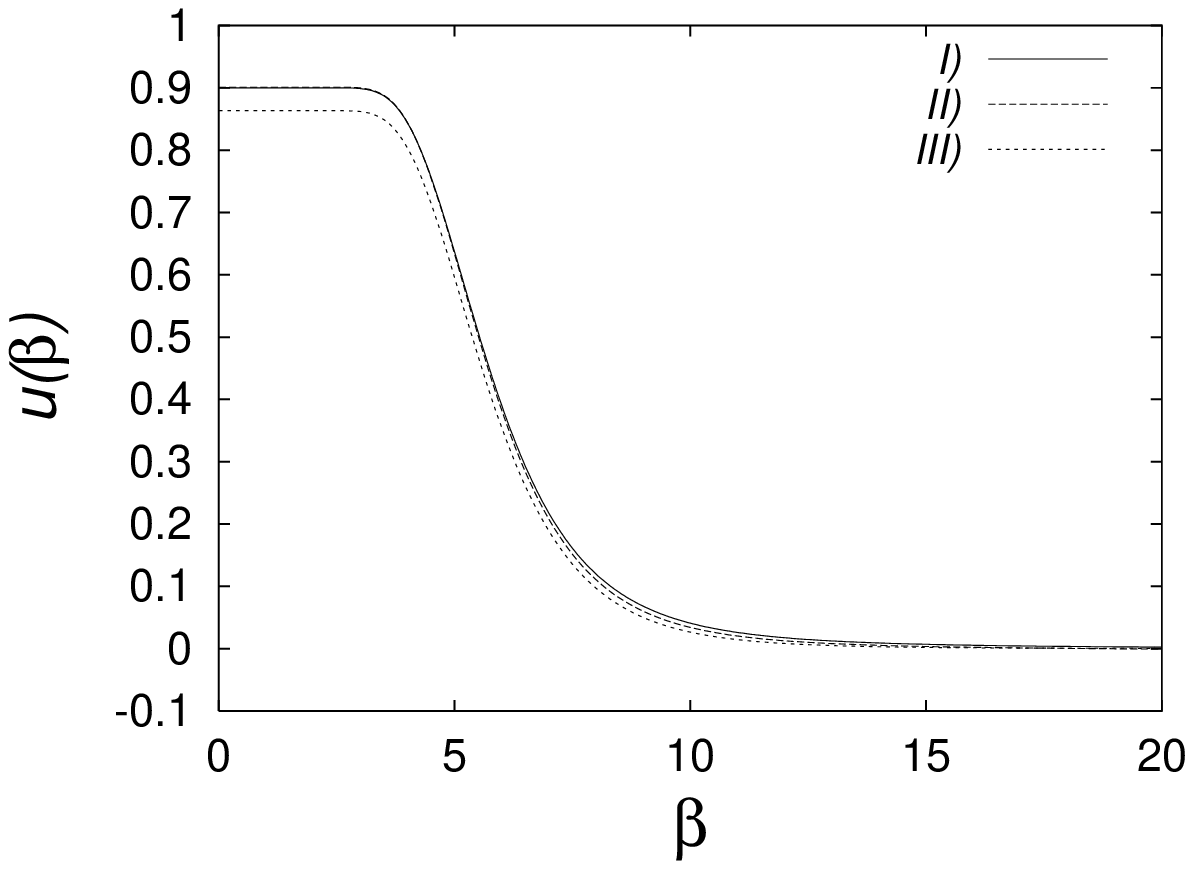,width=8cm} 
\end{center} 
\caption{a) $P_{0}(\beta)$; b) $U(\beta)$. Numerical values of the parameters 
are: $L=10$, $N=400$.} 
\label{fig:pub} 
\end{figure}

The following three cases are studied numerically: 

\noindent 1. We fix the maximal generation, $L$, and the number of jumps, $N$, 
and plot $P_{0}$ and $U$ as the functions of the inverse temperature $\beta$ 
-- see figs.\ref{fig:pub}a,b for the following numerical values of $L$ and 
$N$: $L=10$, $N=400$. 

\begin{figure}[ht] 
\begin{center} 
\epsfig{file=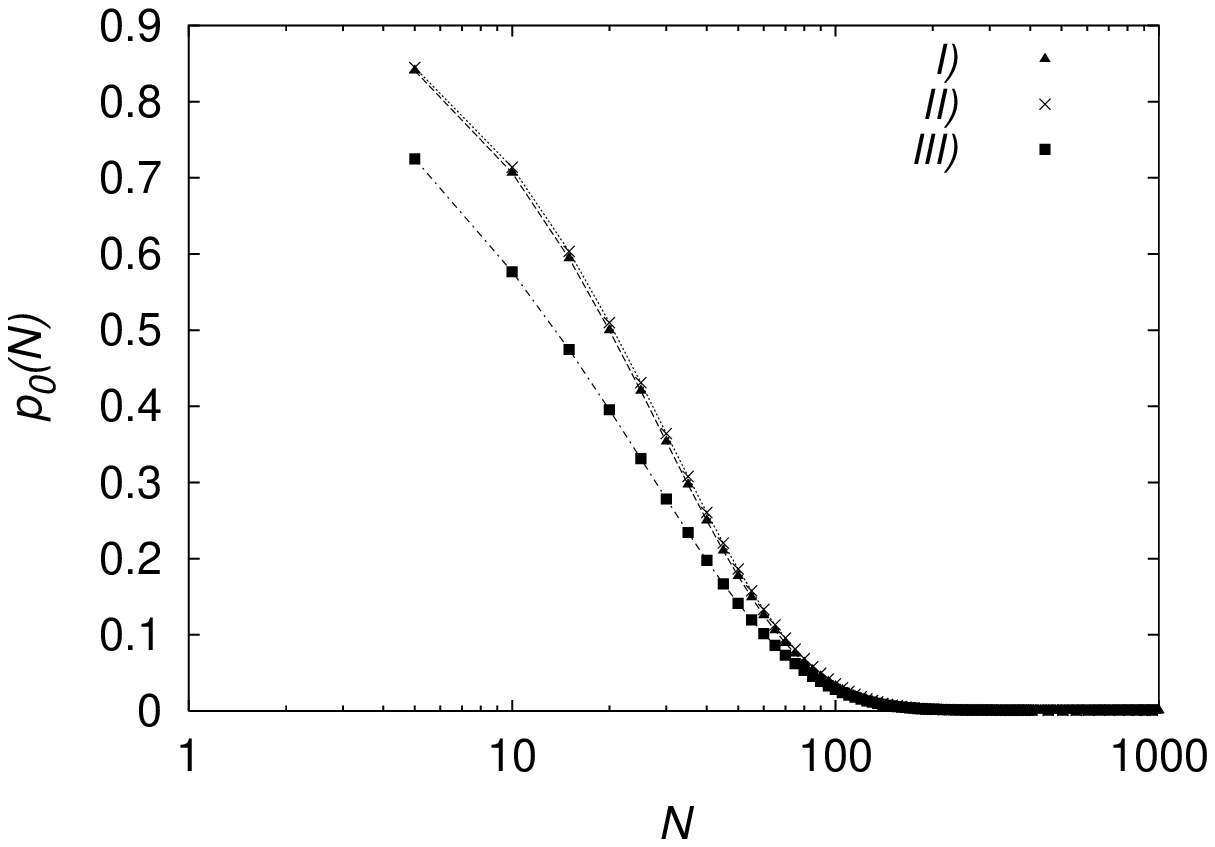,width=8cm} 
\epsfig{file=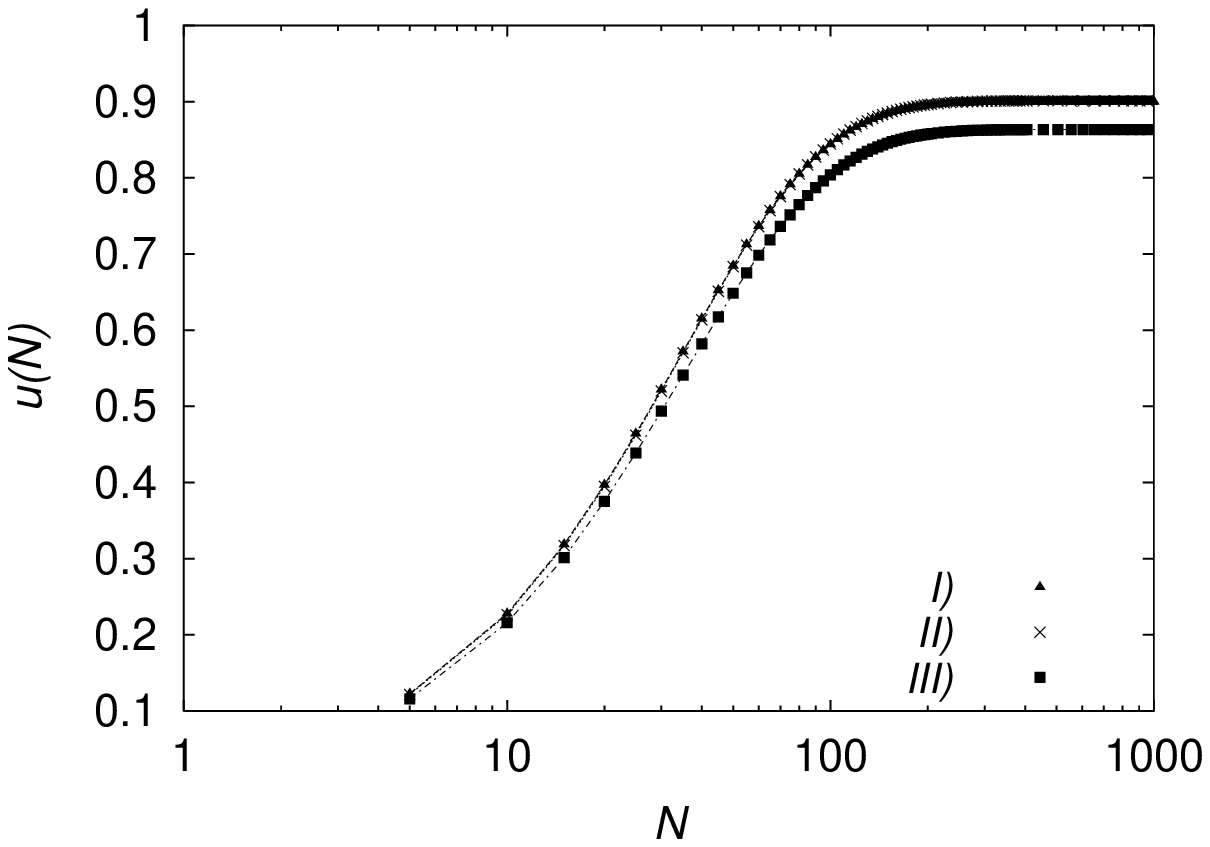,width=8cm} 
\end{center} 
\caption{a) $P_{0}(N)$; b) $U(N)$. Numerical values of the parameters are 
$L=10$, $\beta=4$.} 
\label{fig:pun} 
\end{figure}

\noindent 2. We fix the maximal generation, $L$, and the inverse temperature, 
$\beta$, and plot $P_{0}$ and $U$ as the functions of the number of steps $N$ 
-- see figs.\ref{fig:pun}a,b  for the numerical values of $L$ and $\beta$: 
$L=10$, $\beta=4$. 

\begin{figure}[ht] 
\begin{center} 
\epsfig{file=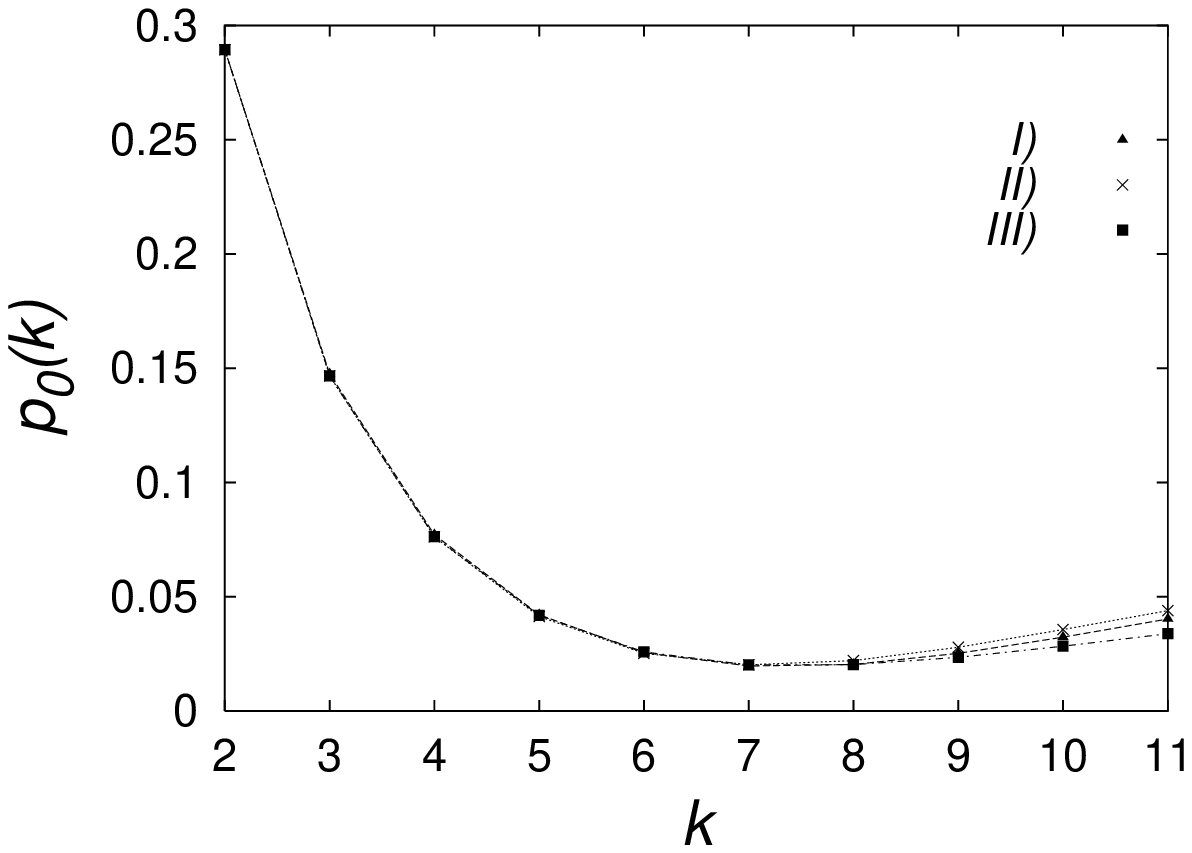,width=8cm} 
\epsfig{file=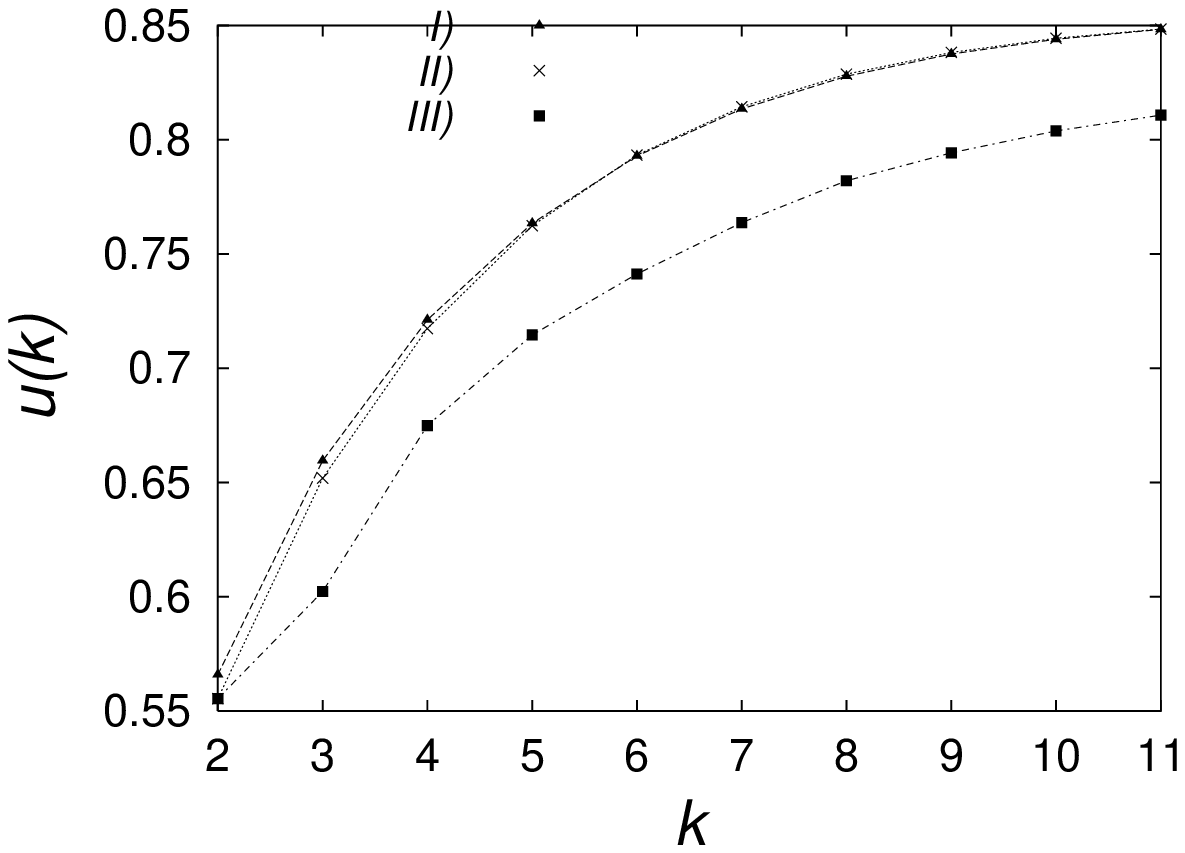,width=8cm} 
\end{center} 
\caption{a) $P_{0}(L)$; b) $U(L)$. Numerical values of the parameters are 
$N=400$, $\beta=1$.} 
\label{fig:puk} 
\end{figure}

\noindent 3. We fix number of jumps, $N$, and the inverse temperature, 
$\beta$, and plot $P_{0}$ and $U$ as the functions of the maximal generation 
$L$ -- see figs.\ref{fig:puk}a,b for the following numerical values of $N$ and 
$\beta$: $N=400$, $\beta=1$. 

\section{Discussion} 
\label{sect:3}

The numerical solution of the master equations (\ref{eq:14}) and (\ref{eq:15}) 
shown in figs.\ref{fig:pub}--\ref{fig:puk} allow us to conclude that we arrive 
to the same results for the return probability $P_0$ and the mean height of 
the barriers $U$ for random walks in the ultrametric spaces independent on 
which model we consider: either the ultrametric structure of the barriers is 
defined on the boundary of the Cayley tree (case I), or  the heights of these 
barriers are given by the properly normalized Dedekind function (cases II and 
III). 

Our results are obtained under the condition that the heights of the 
ultrametric barriers in all models (I,II and III) have {\it linear dependence} 
on the number of Cayley tree generation. Recall that the construction of our 
ultrametric potential (\ref{eq:defbar}) on the basis of Dedekind modular 
function is conditioned only by this demand. The current choice of the 
potential is stipulated basically for demonstrational aims. Namely, we have 
shown that it is possible to adjust the ultrametric structure of the 
continuous metric space ${\cal H}$ to the structure of the standard Cayley 
tree in such a way that we can mimic the main statistical properties of the 
random walk at the boundary of the Cayley tree by the diffusion in ${\cal H}$. 

The key ingredient of our construction is contained in the replacement of the 
RSB scheme (\ref{eq:Q}) by the new MRSB scheme (\ref{eq:U}). Despite yet the 
corresponding master equation (\ref{eq:15}) allows only numerical treatment, 
the proposed construction, to our opinion, has some advantages with respect to 
the RSB scheme resulting from: i) the self-duality (see eq.(\ref{eq:9a})), and 
ii) the continuity of MRSB. 

We permit ourselves to touch below both these properties (i) and (ii) in 
connection with the possible simplification of the solution of the continuous 
analog of the master equation (\ref{eq:16n}) (see part \ref{sect:mrsb} below), 
and with the speculation about the geometry of $n\to 0$ limit (see part 
\ref{sect:n0} below). We clearly realize that the last subject is one of the 
mostly intriguing question is the statistical physics of disordered systems 
and our conjecture might be criticized from different point of view, however 
we believe that our geometric construction might stimulate the readers to fill 
the standard mode of thinking about the replicas by some fresh geometric 
sense. 

\subsection{The continuous MRSB model} 
\label{sect:mrsb} 

Taking advantage of our construction of ultrametric system of barriers 
described by the potential $U_{MRSB}(x_{m})$ (see (\ref{eq:defbar})), we can 
replace the master equation (\ref{eq:14}) by its analog based on the 
properties of the Dedekind function. The distribution function $W_N(x)$ is 
described by (\ref{eq:15}) where $x_m$ is the rational point on the interval 
$[x',x]$ with the smallest denominator (there is only one such point). As we 
have already seen, the highest barrier $U_{MRSB}(x_m)$ on the interval 
$[x',x]$ is located at the point $x_m$. The number of barriers and their 
heights are controlled simultaneously by the parameter $y$. As $y\to 0$, more 
and more new barriers appear. Gradually they become higher and narrower 
preserving the ultrametric structure. In this section we are interested in the 
limit $y\to 0$, where the hierarchical structure of barriers is highly 
developed---see, for example fig.\ref{fig:barvid}. Simplification of 
eq.(\ref{eq:15}) is allowed by using eq.(\ref{eq:9a}). Making in 
eq.(\ref{eq:15}) the substitution which maps $y$ to $\frac{1}{y}$ we get: 
\be 
U_{MRSB}(x_m,y,L^*)\equiv 1+\frac{1}{c\,L^*(y)} \ln \frac{\ln f(x_m,y)}{\ln 
f(\frac{1}{2},y)}\bigg|_{y\to 0}\simeq 1+\frac{2}{c\,L^*(y)}\ln\frac{2}{q} 
\label{eq:y0} 
\ee 
where $L^*(y)$ is determined again using the relation (\ref{eq:y_star}). The 
equation (\ref{eq:y0}) is valid only for $\{p,q,s,r\}\in {\mathbb Z}$ such 
that $ps-qr=1$ (where $r$ might be an arbitrary integer). The point 
$x_m=\left\{\frac{p}{q} \right\}$ and its "dual image" 
$x'_m=\left\{\frac{s}{q}\right\}$ lie in the unit interval $[0,1]$. 
Substituting (\ref{eq:y0}) into (\ref{eq:15}), we have: 
\be 
\left\{\begin{array}{l} \disp W_{N+1}(x)=\int_0^1 dx'\, e^{-\beta} 
\left(\frac{2}{q}\right)^{2\beta/(c\,L^*(y))} \; W_N(x') \\ \disp 
W_{N=0}(x)=\delta_{x,x_0} \end{array}\right. 
\label{eq:15a} 
\ee 
Eq.(\ref{eq:15a}) has rich number theoretic structure. In fact, the 
information of the position of the highest barrier on the interval $[x',x]$ is 
hidden in the integer $q$. Let us recall that the value of $q$ should satisfy 
the following condition: {\it $q$ is the lowest denominator of the rational 
point $x_m=\frac{p}{q}$, $x_m\in [x',x]$}. As one sees, the value $L^*(y)$ 
governs the amplitude of the potential, while $q$ controls the maximal 
"resolution" and, hence, the total number of barriers. 

\subsection{Speculations about continuous number of Cayley tree generations 
$L$ and replica $n\to 0$ limit} 
\label{sect:n0} 

\begin{figure}[ht] 
\begin{center} 
\epsfig{file=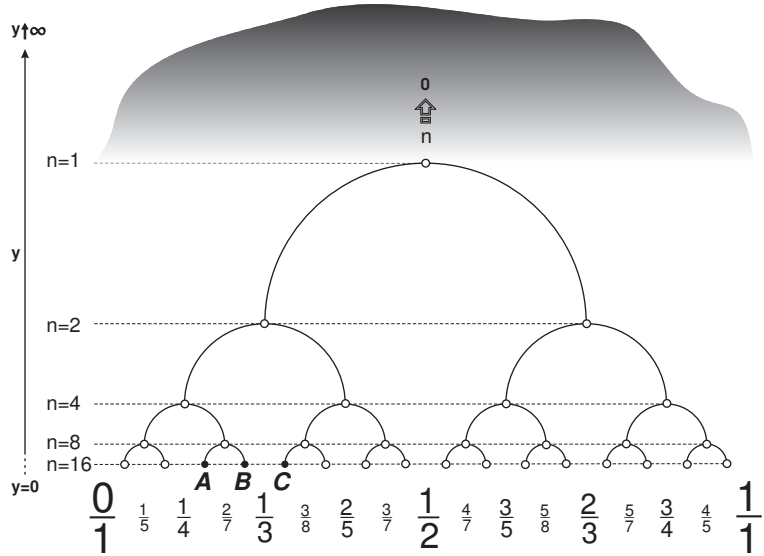,width=10cm} 
\end{center} 
\caption{The 3--branching Cayley tree. The barriers are located at the 
rational points ordered in the Farey sequence. The the replica limit $n\to 0$ 
is interpreted as the absence of states (vertices of the tree) when 
$y\to\infty$.} 
\label{fig:3b} 
\end{figure}

Despite of many advantages of the RSB scheme, it contains the unavoidable in 
the replica formalism mystery of the $n\to 0$ limit, where $n=2^L$ is the 
number of replicas---see fig.\ref{fig:3b}. 

One of the merits of our construction based on the Dedekind $\eta$--function 
consists in the possibility to change continuously the number of ultrametric 
barriers and their heights by varying the parameter $y$, and hence to change 
continuously the number of states $n$ of the RSB matrix. Let us remind that 
the standard 3--branching Cayley tree being isometrically embedded in the 
space ${\cal H}=\{z|\,{\rm Im}\,z>0\}$ becomes a "continuous" structure: one 
can define the smoothed neighborhood of maxima of the function $f(z)$ in the 
half--plane $y={\rm Im}\,z>0$---see fig.\ref{fig:1}. For instance, our 
construction allows us to increase continuously the number of states $n$ from, 
say, $n=1$ to $n=2$ changing correspondingly the parameter $y$ in the interval 
$[y_{n=1}, y_{n=2}]$, where $y_{n=1}=\frac{\sqrt{3}}{6}$ and 
$y_{n=2}=\frac{\sqrt{3}}{14}$.

Our geometric construction based on the isometric embedding of the Cayley tree 
into the complex plane permits us to consider also the opposite limit---the 
case when $n$ is formally less than 1. This case means the absence of any 
states in the RSB matrix ($n=0$). As one sees from figs.\ref{fig:1} and 
\ref{fig:2}, the function $f(z)$ has no longer local maxima above the value 
$y\equiv{\rm Im}\,z>\frac{\sqrt{3}}{2}$. As soon as we have identified the 
local maxima of the function $f(z)$ with the vertices (states) of the Cayley 
tree, we conclude that the absence of any states of RSB matrix means the 
absence of local maxima of the function $f(z)$. Thus, the limit $n\to 0$ we 
interpret as $y\to\infty$. This claim is based on the behavior of the function 
$f(z)$ in the upper half--plane. The correspondence of the limits $n\to 0$ and 
$y\to\infty$ is schematically illustrated in the fig.\ref{fig:3b}. The 
geometrical interpretation of number of replica states in the region $0<n<1$ 
makes the physical statement about the "analytic continuation $n\to 0$" more 
formal. Actually, let us recall the definition of the hyperbolic distance 
${\cal L}$ in ${\cal H}(z|{\rm Im}\,z>0)$ between two points $z_0=(x_0,y_0)$ 
and $z=(x,y)$: 
\be
\cosh {\cal L}=1+\frac{(x-x_0)^2+(y-y_0)^2}{y_0\,y} 
\label{eq:hyp} 
\ee 
As we have seen, the hyperbolic distance ${\cal L}$ is the continuous analog 
of the number of Cayley tree generations, $L$, what permits us to define the 
number of replica states $n$ as 
\be 
n=e^{{\rm const}\,{\cal L}} 
\label{eq:n0} 
\ee 
As $y\to 0$, we can replace $\cosh{\cal L}$ with the leading exponential term. 
This way we arrive at the standard relation $e^{\cal L}=\frac{2y}{y_0}$. 
Hence, 
\be 
\frac{e^{\cal L}}{2} + \frac{e^{-\cal L}}{2} \hspace{-0.64cm} 
\bigg\backslash \;\; =\frac{(x-x_0)^2+y_0^2}{y_0\,y}
\label{eq:l+}
\ee 
neglecting the second term in the left-hand side of (\ref{eq:l+}), we get
$$
{\cal L}=-\ln y+ c_1 \qquad (y\to 0)
$$
where $c_1=\ln \frac{2((x-x_0)^2+y_0^2)}{y_0}$. 

However if $y\to\infty$, eq.(\ref{eq:hyp}) gives 
\be
\frac{e^{\cal L}}{2} \hspace{-0.5cm} 
\bigg\backslash\; + \frac{e^{-\cal L}}{2}=\frac{y}{y_0}
\label{eq:l-}
\ee
where according to the physical condition $n\to 0$ and eq.(\ref{eq:n0}), we 
have to take another branch of the function ${\rm arccosh}(...)$, 
corresponding to the negative  values of ${\cal L}$. This leads to the 
following definition of ${\cal L}$: 
\be 
{\cal L}=-\ln y + c_2 \qquad (y\to \infty)
\label{eq:def_L} 
\ee 
where $c_2=\ln \frac{y_0}{2}$. As $y\to \infty$, the hyperbolic distance 
${\cal L}$, defined according to (\ref{eq:def_L}) tends to $-\infty$. Equation 
(\ref{eq:def_L}) is consistent with the definition (\ref{eq:n0}) of number of 
replicas $n$ in the region $0<n<1$. 

It would be very desirable to check how our conjecture works in the models 
possessing the RSB symmetry of the order parameter. We expect that our 
construction might be useful for the investigation of aging phenomena 
considered from the point of view of the diffusion in the whole ultrametric 
space ${\cal H}(z|{\rm Im}\, z>0)$.

\acknowledgments 
The main part of this work has been accomplished owing to the hospitality of 
the laboratory LIFR-MIIP (CNRS, France and Independent University, Moscow); 
O.V. thanks the laboratory LPTMS (Universit\'e Paris Sud, Orsay) for the 
hearty welcome.

\end{document}